\documentclass[10pt,superscriptaddress,nofootinbib]{revtex4-2}
\usepackage{amsmath,amssymb,amsfonts,epsfig,graphicx,euscript}
\usepackage{graphicx}
\usepackage{xcolor}
\usepackage{lipsum}
\usepackage{amsmath,amssymb,amsfonts,amsthm}

\theoremstyle{plain}

\theoremstyle{definition}

\theoremstyle{remark}
\usepackage{braket}
\usepackage[font=small]{caption}
\usepackage[hidelinks]{hyperref}
\usepackage{url}

\begin{document}
	\begin{center}
		{ \Large{\textbf{Holographic entanglement entropy and mutual information in deformed field theories at finite temperature}}}
		\vspace*{1.2cm}
		\begin{center}
			{\bf Hajar Ebrahim$^{a,1}$, Monireh Ahmadpour$^{a,2}$
			}\\%
			\vspace*{0.3cm}
			{\it {${}^a$Department of Physics, University of Tehran, North Karegar Avenue, Tehran 14395-547, Iran}}
			\vskip 5mm
			{\it {${}^1$hebrahim@ut.ac.ir}, {${}^2$ahmadpour.monireh@ut.ac.ir}}
			\vspace*{0.5cm}
		\end{center}
    \end{center}
\bigskip
\begin{center}
	Abstract
\end{center}
\hspace{\parindent}
In this paper we have studied the effect of deformation and temperature on holographic entanglement entropy and mutual information between two subsystems in a deformed field theory at finite temperature. The $T{\overline{T}}$ deformation operator in field theory introduces a cut-off in the dual gravity background. We define different regimes of parameters and calculate the entanglement entropy and mutual information analytically. We observe that temperature and deformation parameter have similar effects on entanglement entropy and mutual information. They both lead to decrease in mutual information or somehow the correlation between the two subsystems. Interestingly we observe the emergence of correction terms in entanglement entropy and mutual information that are universal and do not depend on the scale of the non-local operator. 

\vspace*{0.7cm}
\tableofcontents
	\flushbottom
	\section{Introduction and results}
Recent developments in the connections between quantum gravity and quantum information in the holographic context \cite{Maldacena:1997re,Gubser:1998bc,Witten:1998qj} has developed a lot of interest in this subject. These developments cover a vast range of topics from entanglement entropy to tensor networks and complexity. The question on these connections can be generalized to situations in holography where a hard radial cut-off has been introduced in the bulk AdS gravity. A useful approach in this regard claims that such gravity theory is dual to a  deformed field theory. The field theory is deformed with some operator so that the hard cut-off, a finite cut-off with Dirichlet boundary condition in the bulk, is reproduced. 	

Discussions on deformations in field theories go back to $T\bar{T}$ deformation introduced in 2-dimensional quantum field theories as a solvable irrelevant deformation \cite{zamolodchikov:2004expectation,smirnov:2017space,Cavaglia:2016oda}, where $T$ is the holomorphic component of the 2-dimensional stress-energy tensor. Later on, such deformation was generalized to other bilinear operators in conserved currents of the theory, operators such as $J\bar{T}$ with $J$ a conserved holomorphic $U(1)$ current \cite{Guica:2017lia,Aharony:2018ics,Nakayama:2018ujt} and to non-relativistic systems \cite{cardy:2018t,chen:2021note} where one does not assume Lorentz invariance. A review on solvable irrelevant deformations of 2-dim field theory can be found in \cite{Jiang_2021}. 

Despite some discussions on the unitarity of the theory in presence of $T\bar{T}$ deformation, a proposal has been made for a holographic dual of an explicitly deformed CFT which produces the bulk physics with a sharp cut-off in $AdS_3$ \cite{McGough:2016lol}. This idea was further developed in \cite{Kraus:2018xrn,Shyam:2017znq,Cottrell:2018skz,Coleman:2020jte}. Inspired by such set-up the generalizations have been made to higher dimensions to understand holography with a hard radial cut-off in $AdS_{d+1}$ \cite{Taylor:2018xcy, Hartman:2018tkw}. The dual effective field theory appears as a deformation of large N conformal field theory with an irrelevant operator, called $X$ which is bilinear in components of the stress tensor and transforms as scalar under Lorentz transformations. This operator reduces to $T\bar{T}$ deformation in 2-dimensional CFT. 

This proposal is very intuitive and due to its simplicity one can apply it to compute various quantities in the bulk and compare with the boundary results. This helps us better understand the effect of $T\bar{T}$ deformation on physical quantities in deformed field theory. Two notable examples with great interest are correlation functions of local operators and quantum information measures such as Renyi and entanglement entropies. Correlation functions of $T\bar{T}$ deformed CFTs were first studied in the large c limit in \cite{Aharony:2018vux}. And the $T\bar{T}$ deformed entanglement entropy was first discussed in \cite{donnelly:2018entanglement} and further developed in \cite{donnelly:2018entanglement,ota:2019comments,Park:2018snf,Asrat:2020uib,banerjee:2019entanglement,grieninger:2019entanglement,donnelly:2020quantum}. For example, in \cite{banerjee:2019entanglement} the entanglement entropy at zero temperature has been calculated using replica trick in deformed field theory and has been compared with the entanglement entropy calculated using holographic dual theory by Ryu-Takayanagi formula \cite{r01}. It has been shown that the entanglement entropy calculated in deformed field theory matches perfectly with the universal terms of the holographic entanglement entropy. In recent years, it has been shown that Ryu-Takayanagi formula is valid for any general holographic theories that obey a GKPW-like dictionary of AdS/CFT correspondence with Dirichlet boundary conditions which are imposed on a finite cut-off surface in the bulk \cite{Murdia:2019fax}.
	
Along with the holographic entanglement entropy another quantum information measure to investigate in the presence of the $T\bar{T}$ deformation is the holographic mutual information (HMI). It is defined as the linear combination of the entanglement entropies, $I(A:B)=S(A) +S(B) -S(A\cup B)$, where S denotes the entanglement entropy of the disjoint regions A and B. It is a semi-definite quantity which measures the total classical and quantum correlations  between the two subsystems A and B \cite{Headrick:2010zt,r05}. It is also a scheme-independent, positive quantity which is proportional to the area of the subsystem at finite temperature. 

	In this paper we try to answer this question that how the deformation (hard cut-off in dual gravity background) will affect the entanglement entropy and mutual information in a deformed field theory at finite temperature, using its holographic recipe\footnote{While we were in the final stages of this work the paper \cite{Jeong:2022jmp} appeared on arxiv which has calculated the entanglement entropy in hyperscaling violating geometries with cut-off and at finite temperature. A few results on entanglement entropy might be similar.}. We consider the asymptotically AdS black brane geometry in general $(d+1)$ dimensions which is dual to the desired field theory mentioned earlier. We discuss such question using the analytic results we obtain in different limits of finite temperature and non-zero deformation parameter. The analytic technique we use here is introduced in \cite{Fischler:2012ca}. In certain limits we can compare our results with the known results in literature such as \cite{Fischler:2012ca,r06} and  \cite{Khoeini-Moghaddam:2020ymm} . In \cite{Fischler:2012ca,r06} analytic evaluation of the entanglement entropy and mutual information for subsystems at non-zero temperature has been done. In \cite{Khoeini-Moghaddam:2020ymm} the authors have discussed the effect of cut-off on entanglement entropy and mutual information in field theory at zero temperature. The results of our paper can be summarized as:

\begin{itemize}
\item We consider different regimes of parameters in field theory for the entanglement entropy given by high, low temperature $T$ and also small, large deformation denoted by deformation parameter $\tilde{\lambda}$. The corrections to the entanglement entropy for a subsystem of scale $\ell$ in the deformed field theory at non-zero temperature has been calculated up to order $2d$ in parameters. In general, for different limits of parameters, the corrections due to both temperature and deformation parameter decrease the entanglement entropy. These corrections include three dimensionless expansion parameters, one due to temperature which is $T\ell$, one due to deformation parameter $\tilde{\lambda}/\ell$ and the one that emerges from the calculations is the product of temperature and deformation parameter $\tilde{\lambda} T$. The last one only depends on field theory parameters but the first two ones know about the non-local operator which is calculated.  

\item We observe that in the low temperature and small deformation regime, subsection \ref{low small}, all the corrections to the entanglement entropy follow area law behaviour. From entanglement point of view, the corrections due to temperature and deformation are similar as if the deformation acts the same way as temperature in deformed field theory. An interesting result here is that the divergent piece of the entanglement entropy in the undeformed theory gets corrected only by a series in $\tilde{\lambda} T$ as can be seen in the first line of (\ref{firstlimit}). Such a correction exactly repeats itself in the high temperature and small deformation limit as in second line of (\ref{thirdlimit}). 

\item In the low temperature and high deformation limit, subsection \ref{low large}, the behaviour of the entanglement entropy is somehow different. The leading order term in the entanglement entropy, relation (\ref{secondlimit}), behaves as volume law which is divergent if the deformation parameter goes to zero. Therefore in the large deformation regime the dominant contribution to the entanglement entropy is the thermal entropy. Since we are probing the area near the cut-off the correction terms due to temperature only appear in the series with expansion parameter which is a dimensionless product of temperature and deformation parameter. The correction terms due to deformation or cut-off get a universal form, independent of the dimension of field theory. 

\item In the high temperature and small deformation limit, subsection \ref{high small}, we are probing the area near the horizon. Therefore the corrections due to the deformation in entanglement entropy appears only as a dimensionless product of temperature and deformation parameter as in (\ref{thirdlimit}). The leading order correction due to temperature is proportional to the volume of the subsystem which shows that the entanglement entropy behaves as thermal entropy. Interestingly in this limit, except for the leading temperature correction term, all the other corrections are series in the dimensionless product of temperature and deformation parameter. Therefore the correction terms do not distinguish between temperature and deformation parameter in this limit.  

\item The other quantity that we calculate is the mutual information in different regimes of parameters. We observe that mutual information follows an area law behaviour as expected and deformation parameter does not change it. We also see that increasing the temperature and deformation parameter leads to the disentanglement of the subsystems at smaller separation distances. In the limit of small deformation, subsection \ref{subsmalldef}, mutual information diverges as separation distance goes to zero even for finite temperature and finite deformation parameter. The correction terms in the mutual information consists of individual terms due to temperature and due to deformation parameter in addition to terms involving the dimensionless combination of the two. We plot phase transition diagrams with respect to different parameters of the theory. The constant value at which the phase transition between zero and non-zero mutual information happens depends on dimension of the field theory, temperature, scale of the subsystems and deformation parameter. The plots show that at smaller value for the deformation parameter the separation at which the phase transition happens is less dependent on temperature than larger values of it. 

\item Mutual information in the large deformation regime, subsection \ref{sublargedef}, differs from the result in the small deformation regime because the mutual information does not diverge in the zero separation limit. This is due to the dominant effect of the deformation parameter compared to the temperature. In fact in the large deformation regime temperature does not affect the mutual information independently of the deformation parameter. The only effect of temperature is through the dimensionless product of it with the deformation parameter. In this limit we plot the phase diagrams too and the results are not different from the previous limit.  

\item In the intermediate deformation regime, subsection \ref{subinterdef}, zero separation distance does not lead to divergent mutual information. This result is similar to the large deformation one. Also similar to small deformation regime there are individual expansions in temperature and deformation parameter.
\end{itemize}
More details on the results have been discussed throughout the paper. The organization of the paper is as follows. In the first two sections we introduce the background and review the $T\bar{T}$ deformation in field theory dual to the asymptotically $AdS$ background. In section \ref{sec03} and \ref{sec001} we discuss the entanglement entropy in deformed field theory at finite temperature in different regimes of parameters. And finally in section \ref{sec04} we obtain mutual information in different limits and discuss how it changes by varying the parameters.

\section{$T\bar{T}$ deformation of $CFT_{d}$}\label{sec02}	
In this section we briefly review how the cut-off in $AdS_{d+1}$ background can be related to analogous $T\bar{T}$ deformation in d-dimensional $CFT$. We also obtain what the dictionary between parameters in deformed field theory and gravity dual is in our set-up. For more details we refer the interested reader to \cite{Taylor:2018xcy, Hartman:2018tkw}. We assume that the classical action of CFT has been deformed by an operator named $X$ which is considered to be the generalization of 2-dimensional $T\bar{T}$ deformation in d-dimensional CFT \cite{Taylor:2018xcy}
\begin{equation}\label{effective}
S_{EFT}=S_{CFT}+\lambda~\int {d^dx \sqrt{\gamma}~X}~,
\end{equation}
where $\gamma_{ij}$ is the boundary theory metric. As it is proposed by \cite{McGough:2016lol,Taylor:2018xcy, Hartman:2018tkw} $T\bar{T}$ deformation for $\lambda>0$ (in our convention) is dual to a cut-off AdS geometry. The deformation operator is considered as a solvable irrelevant deformation and has mass dimension $2d$~. The deformation parameter $\lambda$, which shows the strength of deformation, is dimensionful with mass dimension $\Delta_\lambda = -d$. Therefore invariance of the effective action for such theory which includes only one dimensionful parameter implies
\begin{equation}\label{onshell}
  \lambda~\frac{d W}{d\lambda} =\frac{1}{\Delta_{\lambda}} \int {d^dx \sqrt{\gamma}~ \langle T \rangle}~,
\end{equation}
where $T = \gamma^{ij} T_{ij}$ is the trace of the field theory stress-energy tensor and $W$ is the on-shell effective action on the cut-off hypersurface at $z=z_c$. Comparing this result (\ref{onshell}) with the derivative of the general form of effective action in (\ref{effective}) implies that
\begin{equation}\label{flow_eq}
  \langle T \rangle = - d \lambda~ X~.
\end{equation}
Therefore the deformation operator should be equal to the trace of field theory stress tensor up to a coefficient proportional to the deformation parameter $\lambda$.

Now we would like to see what the form of deformation operator $X$ is in terms of field theory stress-energy tensor and bulk parameters. Let's assume that the bulk metric has the general form
\begin{equation}
ds^2 = g_{\mu\nu} dx^\mu dx^\nu = \frac{R^2}{z^2} \left(dz^2 + \gamma_{ij}(z,x) dx^i dx^j\right)~,
\end{equation}
where $\gamma_{ij}(z,x)$ is the boundary theory metric. We assume that the intrinsic curvature of the constant $z=z_c$ hypersarface or boundary theory is zero. Therefore for the bulk action in the presence of the counterterm and Gibbons-Hawking term we have
\begin{align}
  S_{bulk} &= S_{EH}+S_{GHY}+S_{ct}\nonumber\\
           &= \frac{1}{16 \pi G_{N}^{d+1}}\int {d^{d+1}x \sqrt{g}(\mathcal{R}- \frac{d(d-1)}{R^2})}
           -\frac{1}{8\pi G_{N}^{d+1}}\int {d^{d}x \sqrt{h}K}
           + \frac{1}{8\pi G_{N}^{d+1}}\int {d^{d}x \sqrt{h}(\frac{d-1}{R})}~,
\end{align}
where $\mathcal{R}$ is the bulk curvature and $h_{ij}$ is the metric of the time-like cut-off hypersurface at $z=z_c$ where field theory lives. If we call the normal vector to the time-like hypersurface $n^\mu$ the metric $h_{\mu\nu}$ is $g_{\mu\nu}-n_\mu n_\nu$. For the metric we have chosen here, the only non-zero components of the hypersurface metric are $h_{ij}=\frac{R^2}{z_c^2}\gamma_{ij}(z_c,x)$. We also have $K_{ij} = h_i^m \nabla_m n_j$ which is the extrinsic curvature of the cut-off hypersurface.  Note that we have assumed the field theory lives on a flat manifold and therefore the curvature counterterms vanish and the intrinsic curvature is zero. The interested reader can check the extra terms in \cite{deHaro:2000vlm}. We have also assumed that there is no action for matter fields in our set-up. Therefore the variation of the action with respect to the metric on the cut-off hypersurface $h_{ij}$ leads to Brown-Youk stress-energy tensor which we denote it by $\tilde{T}_{ij}$ and is
 \begin{align}
 \tilde{T}_{ij}=\frac{1}{8\pi G_{N}^{d+1}}\left(K_{ij}-K h_{ij}+\frac{d-1}{R}h_{ij}\right)~.
 \end{align}
 Note that using Gauss-Codazzi equation $\tilde{T}_{ij}$ satisfies the following equation
 \begin{equation}\label{GC}
 \tilde{T}_{ij} \tilde{T}^{ij} - \frac{1}{d-1} {\tilde{T}} ^2 = \frac{-1}{4 \pi G_N^{(d+1)} R} \tilde{T}~,
 \end{equation}
 where $\tilde{T}$ is the trace of the cut-off hypersurface stress-energy tensor $\tilde{T} = h^{ij}\tilde{T}_{ij}$. On the other hand, from AdS/CFT we know that the boundary theory stress-energy tensor, which we call it $T_{ij}$ is obtained by the variation of the renormalized bulk or gravity action with respect to the boundary metric $\gamma_{ij}$ \cite{Balasubramanian:1999re}. Therefore we have
 \begin{align}\label{set}
\langle T_{ij} \rangle= \frac{-2}{\sqrt{\gamma}}\frac{\partial S_{gr,ren}}{\partial \gamma^{ij}} = \frac{R^{d-2}}{z_c^{d-2}} \langle \tilde{T}_{ij} \rangle~.
 \end{align}
This is the relation between boundary theory stress-energy tensor and the energy-momentum tensor on the cut-off hypersurface. It has been suggested in \cite{Taylor:2018xcy,Hartman:2018tkw} that in general d dimensions the deformation operator that one should consider has the form 
\begin{equation}\label{TT_operator}
 X=T_{ij}~T^{ij}-\frac{1}{d-1} T^2~,
\end{equation}
which is written in terms of the boundary theory stress-energy tensor. This operator reduces to the $T\bar{T}$ deformation in 2 dimensions. Using (\ref{set}) the operator $X$ can be written in terms of the cut-off hypersurface stress-energy tensor $\tilde{T}_{ij}$ . Therefore we will be able to use relation (\ref{GC}) and write $X$ solely in terms of the trace of the boundary theory stress-energy tensor $T$. Replacing $X$ in equation (\ref{flow_eq}) we can obtain the field theory deformation parameter $\lambda$ in terms of the bulk parameters as
\begin{equation}\label{lambda}
  \lambda = \frac{4 \pi G_N^{(d+1)}}{d~ R^{d-1}} z_c^d~,
\end{equation}
which has length dimension $d$ as expected. This relation shows that changing the deformation parameter $\lambda$ explores different radii of bulk spacetime. Note that to be able to compare our results with the original undeformed conformal field theory we will work in the limit $\lambda \to 0$.

\section{The background geometry}\label{sec01}
	As mentioned in the introduction we are interested in studying entanglement entropy in a deformed field theory at finite temperature. Therefore the background geometry dual to such field theory can be considered as  ($d+1$)-dimensional asymptotically AdS black brane geometry with cut-off at $z=z_{c}$, where field theory lives on flat spacetime. The gravity geometry is
\begin{equation}\label{bulk_metric}
ds^2=\frac{R^2}{z^2}\left( - f(z) dt^2+ d{\bold{x}}^2_{d-1}+\frac{dz^2}{f(z)} \right),
\end{equation}
where $R$ is the AdS radius and $f(z)=1-\frac{z^d}{z_H^d}$ is the blackening factor. By Euclidean continuation of this metric,  the field theory temperature $T$ which is Hawking temperature of the black hole reads
\begin{equation}\label{temperature}
  T=\frac{d}{4\pi z_H}~.
\end{equation}
After a brief introduction on the background and the deformed theory, we are ready to go through the main material of the paper. In the next section we shortly review the entanglement entropy and calculate it in the deformed field theory we discuss in this paper. 

    \section{Holographic entanglement entropy}\label{sec03}
    
	Entanglement entropy is a non-local quantity which measures the amount of information loss in a subsystem. In order to calculate the entanglement entropy one can consider a $CFT_d$ as a system living on a Cauchy surface $\mathcal{C}$ of d-dimensional Lorentzian flat manifold $N_d$~. This surface can be divided into two subregions A and B such that $A\cup B=\mathcal{C}$. The entangling surface $\partial A$, which is the boundary of region A, is a co-dimension 2 hypersurface in $N_d$. Assume that the total Hilbert space of this system can be factorized into direct product of subregion Hilbert spaces as $\mathcal{H}_{tot}=\mathcal{H}_A \bigotimes \mathcal{H}_{B}$. For a $CFT_d$ at zero temperature the pure ground state $\ket{\psi}$ (without any degeneracy) can describe the system completely and its density matrix is
\begin{equation}\label{total_density_matrix}
 \rho_{tot}=\ket{\psi}\bra{\psi}~.
\end{equation}
 The observer who has access only to the subsystem A would describe it with the reduced density matrix $\rho_A=tr_B\rho_{tot}$ where the subsystem B degrees of freedom are traced out.
\begin{figure}[hb]
  \centering
  \includegraphics[scale=0.6]{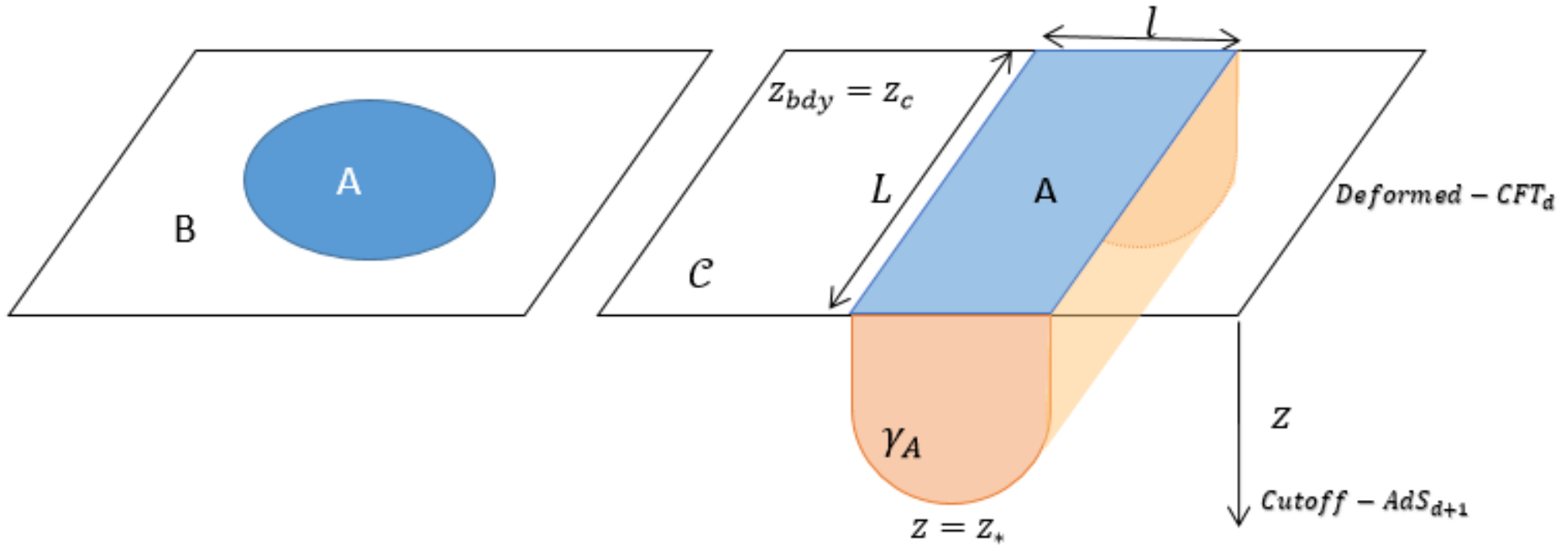}
  \caption{\textbf{Left:} Subregion A and its complement part B($A\cup B=\mathcal{C}$). \textbf{Right:} The schematic diagram of strip A with minimal hypersurface $\gamma_A$ in the bulk anchored on its boundary.}
  \label{fig:entanglingregion}
\end{figure}
The entanglement entropy of the subsystem A is defined as von Neumann entropy of the reduced density matrix $\rho_A$ 
\begin{equation}\label{von_Neumann_entropy}
  S_A\equiv-tr_A(\rho_A~\ln \rho_A)~.
\end{equation}
So the entanglement entropy is the amount of information loss and measures the entanglement between subsystems A and B. For a system at non-zero temperature the total density matrix is replaced with thermal density matrix $\rho=e^{-H/T}$, where H is the system total Hamiltonian.

	The holographic prescription for the entanglement entropy in the $(d+1)$-dimensional asymptotically AdS spacetime $\mathcal{M}_{d+1}$, is given by \cite{r01}
\begin{equation}\label{HEE}
  S(A)=\frac{\mathcal{A}(\gamma_A)}{4G_N^{(d+1)}}~,
\end{equation}
where $\gamma_{\mathcal{A}}$ is a co-dimension 2 extremal surface in the bulk and its area is denoted by $\mathcal{A}(\gamma_{A})$ such that $\partial\gamma_{A}=\partial A$ and $G_N^{(d+1)}$ is the $(d+1)$-dimensional Newton's constant \cite{r01,r02}. We call $\gamma_{A}$ the entangling surface. This proposal satisfies the required properties for any entanglement measure in quantum mechanics \cite{Headrick:2007km,r09,Ryu:2006ef,Headrick:2010zt}. For more information and review on the topic and relevant references we refer the interested reader to \cite{Nishioka:2009un}. 

In the following subsection we discuss how to calculate the area of the entangling surface, $\mathcal{A}(\gamma_{A})$, in our set-up.

	\subsection{Set-up}
	In order to discuss the entanglement entropy in holographic field theory we consider a strip which has length $\ell$ in one of the spatial directions in field theory and the other directions are assumed to have length $L$  (see figure \ref{fig:entanglingregion}). We call $\ell$ the entangling length. This strip is specified by
\begin{equation}\label{strip}
  x\equiv x^1\in \left[-\frac{l}{2},\frac{l}{2}\right]~,~x^i \in  \left[-\frac{L}{2},\frac{L}{2}\right]~,i=2,...,d-1~.
\end{equation}
We assume $L\rightarrow \infty$ to preserve translational invariance of the system in $x^i$ directions. Therefore the profile of the entangling surface in the bulk is given by $x(z)$. Since the area of the entangling surface is the square root of the determinant of the induced metric on it we obtain 
	\begin{equation}
	\label{area}
	\mathcal{A}=L^{d-2} \int dz~ \left(\frac{R}{z}\right)^{d-1} \left(1-\frac{z^d}{z_H^d}\right)^{-\frac{1}{2}}\sqrt{1+x'^2 \left(1-\frac{z^d}{z_H^d}\right)}~,
	\end{equation}
	where $x'=\frac{dx}{dz}$.
	It is easy to see that the integrand, which we call it Lagrangian, has no explicit dependence on $x$. Therefore using the usual method in Lagrangian (or Hamiltonian) formalism we have 
	\begin{equation}
	{x'}^2= \frac{c^2}{\left(\left(\frac{R}{z}\right)^{2(d-1)}-c^2\right)\left(1-\frac{z^d}{z_H^d}\right)}~,
	\end{equation}
		where $c$ is the constant of motion for the, so-called, field $x$ and $x' = \frac{dx}{dz}$. Since the system is symmetric in x direction, we have the extrema of the minimal entangling surface at $z=z_*$ where $z_*$ satisfies the condition $z'(x)=0$. $z_*$ is shown in figure \ref{fig:entanglingregion}. Therefore we get $c=(\frac{R}{z_*})^{d-1}$ and we can write
		\begin{equation}
		\label{xprime}
		\frac{dx}{dz}=\frac{\left(\frac{z}{z_*}\right)^{d-1}}{\sqrt{1-\left(\frac{z}{z_H}\right)^d}\sqrt{1-\left(\frac{z}{z_*}\right)^{2(d-1)}}}~.
		\end{equation}
		Regarding the definition of $x'$ we can obtain $\ell$ by integrating both sides of equation (\ref{xprime}). The integral over $x$ gives $\frac{\ell}{2}$ which becomes
		\begin{equation}
		\frac{\ell}{2}= z_*  \int^1_{\frac{z_c}{z_*}} du~ \frac{u^{d-1}}{\sqrt{1-u^{2(d-1)}}\sqrt{1-(\frac{z_*}{z_H} u)^d}}~,
		\end{equation}
		where we have redefined $z$ as $z_* u$ and $z_c$ is the radial cut-off. This integral can be solved analytically if we use the binomial expansion, first discussed in \cite{Fischler:2012ca}. This expansion is applicable here since, in the range of integral, $\frac{z_*}{z_H} u$ and $u$ are always smaller than $1$. Therefore we have
		\begin{equation}
		\frac{\ell}{2} = z_*~\sum_{k=0}^{\infty}\frac{\Gamma(k+\frac{1}{2})}{\sqrt{\pi} \Gamma(k+1)}\left(\frac{z_*}{z_H}\right)^{kd}\int^1_{\frac{z_c}{z_*}} du~ \frac{u^{(kd+d-1)}}{\sqrt{1-u^{2(d-1)}}}~,
		\end{equation}
		which gives
		\begin{equation}\label{ell2}
			\begin{aligned}
		\frac{\ell}{2} = z_* \sum_{k=0}^{\infty}\frac{\Gamma(k+\frac{1}{2})}{\Gamma(k+1)} \left(\frac{z_*}{z_H}\right)^{kd}&
		\bigg{[}~\frac{\Gamma\left(\frac{d(k+1)}{2(d-1)}\right)}{(1+kd)\Gamma\left(\frac{1+kd}{2(d-1)}\right)}\\
		&- \frac{1}{\sqrt{\pi} d(k+1)}\left(\frac{z_c}{z_*}\right)^{d(k+1)}~ _2F_1\left(\frac{1}{2},\frac{d(k+1)}{2(d-1)};\frac{d(3+k)-2}{2(d-1)},\left(\frac{z_c}{z_*}\right)^{2(d-1)}\right)\bigg{]}~.
		\end{aligned}
		\end{equation}
		Therefore the entangling length, $\ell$, is given in terms of two expansions with expansion parameters $z_*/z_H$ and $z_c/z_*$. This equation should be solved in order to obtain $z_*$, which is a bulk parameter, in terms of known entangling length $\ell$, which is a field theory scale. To do this analytically we can define different regimes of parameters in field theory such as low and high temperature limit and also small and large deformation parameter limit. This helps us solve the above equation perturbatively. We follow the analogy of  \cite{Fischler:2012ca} and give the detailed discussion in the next section. 
		
		Our next step is to obtain the area of the entangling surface. To do so we substitute equation (\ref{xprime}) into equation (\ref{area}) which leads to
		\begin{equation}
		\mathcal{A} = 2R^{d-1} \frac{L^{d-2}}{z_*^{d-2}} ~\sum_{k=0}^{\infty}\frac{\Gamma(k+\frac{1}{2})}{\sqrt{\pi} \Gamma(k+1)}\left(\frac{z_*}{z_H}\right)^{kd}\int^1_{\frac{z_c}{z_*}} du~ \frac{u^{(kd-d+1)}}{\sqrt{1-u^{2(d-1)}}}~,
		\end{equation}
		where the factor 2 is due to the symmetric structure of the entangling surface in $x$ direction. Doing the integral on the right hand side of the above equation produces the area of the entangling surface which is
        \begin{equation}\label{area2}
		\begin{aligned}
		\mathcal{A} = 2R^{d-1}\left(\frac{L}{z_*}\right)^{d-2} &\sum_{k=0}^{\infty}\frac{\Gamma(k+\frac{1}{2})}{\Gamma(k+1)} \left(\frac{z_*}{z_H}\right)^{kd}\bigg[~\frac{\Gamma\left(\frac{d(k-1)+2}{2(d-1)}\right)}{2(d-1)\Gamma\left(\frac{1+kd}{2(d-1)}\right)}\\
		&- \frac{1}{\sqrt{\pi} (d(k-1)+2)}\left(\frac{z_c}{z_*}\right)^{d(k-1)+2}~ _2F_1\left(\frac{1}{2},\frac{d(k-1)+2}{2(d-1)};\frac{d(k+1)}{2(d-1)},\left(\frac{z_c}{z_*}\right)^{2(d-1)}\right)\bigg]~.
		\end{aligned}
		\end{equation}
	This has similar structure to $\ell$ in equation (\ref{ell2}). $z_*$ in this equation needs to be replaced by the solution obtained from the equation (\ref{ell2}). The final result is the area of the entangling surface in terms of expansions in field theory parameters. The full details is given in the following section. 
		
\section{Entanglement entropy in different limits}\label{sec001}

So far we have obtained the general relation for the area of the entangling surface in the gravity dual to a $T\bar{T}$-deformed field theory at non-zero temperature. We used binomial expansion in relations (\ref{ell2}) and (\ref{area2}) for the area and the entangling length rather than solving them numerically. This gives us analytical view over the final results. 

In order to see more directly the effect of temperature and deformation parameter on entanglement entropy, we introduce different limits where we can expand the relations (\ref{ell2}) and (\ref{area2}). Using the dimensionless combination of parameters in field theory we can consider low and high temperature limits, $T\ell \ll 1$ and $T \ell \gg 1$, respectively and small and large deformation limits, $\frac{\tilde{\lambda}}{\ell} \ll 1$ and $\frac{\tilde{\lambda}}{\ell} \gg 1$, respectively. Note that we have defined $\tilde{\lambda} \equiv \lambda ^{\frac{1}{d}}$. Previously we have pointed out that in order to compare our results with the known results in CFT we assume that $\tilde{\lambda} \to 0$ since the $T\bar{T}$ deformation breaks the conformal symmetry. The large deformation parameter should not be confused with this assumption since large deformation is defined in comparison with another scale in field theory $\ell$, the entangling length or subsystem scale. The first two limits on temperature have been discussed previously in holographic strongly coupled CFTs at non-zero temperature in \cite{Fischler:2012ca,r06}. The other two limits on deformation parameter have been discussed in deformed field theories at zero temperature \cite{Khoeini-Moghaddam:2020ymm}. In this paper we study entanglement entropy and mutual information in holographic deformed field theories at non-zero temperature. We would like to see what the combined effect of temperature and deformation parameter is on entanglement entropy and mutual information. 

As we mentioned previously the gravity dual to the $T\bar{T}$ deformed d-dimensional field theory at non-zero temperature is $AdS_{d+1}$ with cut-off and non-zero horizon. The limits introduced in field theory in the last paragraph can be derived in the gravity dual using the black hole horizon at $z_H$, the bulk cut-off at $z_c$ and the turning point of the entangling surface at $z_*$. Therefore the low and high temperature limits translates into $z_* \ll z_H$ and $z_* \simeq z_H$, respectively. The counterparts of the small and large deformation limits are $z_* \gg z_c$ and $z_* \simeq z_c$, respectively. Therefore, in the infinite series for the entanglement length (\ref{ell2}) and the area (\ref{area2}), the above limits can be applied by the combinations $\frac{z_*}{z_H}$ and $\frac{z_c}{z_*}$ to be smaller than or close to 1.

Combining these limits we have three rather physical classes of expansions which are $z_c \ll z_* \ll z_H$, $z_c \ll z_* \simeq z_H$ and $z_c \simeq z_* \ll z_H$. In deformed field theory side they correspond to low temperature with small and large deformation and high temperature with small deformation. We will discus the results for entanglement entropy in these limits in the following subsections. Note that we do not discuss the limit where $z_c \simeq z_* \simeq z_H$ which tells us that $T \tilde{\lambda} \gg 1$. In the limit $\tilde{\lambda} \to 0$ the temperature of field theory becomes very large and therefore due to the complex high energy modes the duality between these two theories is not reliable \cite{McGough:2016lol}.

\subsection{Low temperature and small deformation limit; \textbf{$z_c \ll z_* \ll z_H$}}
\label{low small}
	The first valid range of parameters for the entanglement entropy corresponds to $T\ell \ll 1$ and $\frac{\tilde{\lambda}}{\ell} \ll 1$. Therefore the expansion parameters $\frac{z_c}{z_*}$ and $\frac{z_*}{z_H}$ are small so that the series in $\ell$ (\ref{ell2}) and the area (\ref{area2}) remain convergent. Using the relation for hypergeometric function expansion (\ref{hypergeometric function}) the area and the entangling length relations become
	       \begin{align}\label{ell3}
	       l &= 2 z_* \sum_{k=0}^{\infty} \frac{1}{1+kd} ~\frac{\Gamma(k+\frac{1}{2})~\Gamma\left(\frac{d(k+1)}{2(d-1)}\right)}{\Gamma(k+1)~\Gamma\left(\frac{1+kd}{2(d-1)}\right)} \left(\frac{z_*}{z_H}\right)^{kd}\nonumber\\
	       &-2 z_* \sum_{k=0}^{\infty} \sum_{n=0}^{\infty} \frac{1}{\pi \left(2 n\left(d-1\right)+d\left(k+1\right)\right)} ~\frac{\Gamma(k+\frac{1}{2}) \Gamma(n+\frac{1}{2})}{\Gamma(k+1)\Gamma(n+1)}~\left(\frac{z_*}{z_H}\right)^{kd}\left(\frac{z_c}{z_*}\right)^{d(k+1)+2n(d-1)}~,
	       \end{align}
	      and
	      \begin{equation}\label{Area of entangling region}
	      \begin{aligned}
	      A &= 2R^{d-1} \left(\frac{L}{z_*}\right)^{d-2} \sum_{k=0}^{\infty} \frac{1}{d(k-1)+2} ~\frac{\Gamma(k+\frac{1}{2})\Gamma\left(\frac{d(k+1)}{2(d-1)}\right)}{\Gamma(k+1)\Gamma\left(\frac{1+kd}{2(d-1)}\right)} \left(\frac{z_*}{z_H}\right)^{kd}\cr
	       &-2R^{d-1} \left(\frac{L}{z_*}\right)^{d-2} \sum_{k=0}^{\infty} \sum_{n=0}^{\infty}
 \frac{1}{\pi \left(2+d\left(k-1\right)+2 n\left(d-1\right)\right)} ~\frac{\Gamma(k+\frac{1}{2}) \Gamma(n+\frac{1}{2})}{\Gamma(k+1)\Gamma(n+1)}~\left(\frac{z_*}{z_H}\right)^{kd}\left(\frac{z_c}{z_*}\right)^{2+d(k-1)+2 n(d-1)}~.
	      \end{aligned}
	      \end{equation}
The goal is to obtain the entanglement entropy in terms of physical parameters in field theory $T$, $\tilde{\lambda}$ and $\ell$. The usual path in this regard is to obtain $z_*$ in terms of $\ell$ perturbatively and substitute it in the area relation. Since there are two small expansion parameters in (\ref{ell3}) we contract to keep the terms up to order $2d$ in both parameters together. Doing so $z_*$ becomes
	       \begin{equation}
           \begin{aligned}
	       \label{zstar}
	       z_* = \frac{l}{2 c_0}&\Bigg{\{} 1 + \frac{2}{d} (2 c_0)^{d-1} \left(\frac{z_c}{l}\right)^d-\frac{4}{d^2}\left(d-1\right)(2c_0)^{2(d-1)}
\left(\frac{z_c}{l}\right)^{2d}\\
&-\left(\frac{c_1}{(2 c_0)^{d+1}}+ \frac{c_1}{d {c_0}^2} \left(\frac{z_c}{l}\right)^d -\frac{(2 c_0)^{d-1}}{2d}\bigg(1-\frac{\left(3 d^2-7 d+6\right)c_1}{d {c_0}^2}\bigg)\left(\frac{z_c}{l}\right)^{2d}\right) \left(\frac{l}{z_H}\right)^d\\
&+\Bigg(\frac{2 {c_1}^2 \left(1+d\right) - 3 c_0 c_2}{2 (2 c_0)^{2 (d+1)}} +\frac{{c_1}^2 \left(3 d^2+7 d+6\right) - 6\left(1+d\right) c_0 c_2}{d(2 c_0)^{d + 3}}~\left(\frac{z_c}{l}\right)^{d}\\
&+\frac{1}{8 d^2 {c_0}^4}\left(d\left(d-2\right){c_0}^2 c_1
+ 2\left(5 d^2+6\right){c_1}^2-3\left(2 d^2 + d + 3\right)c_0 c_2 \right)~\left(\frac{z_c}{l}\right)^{2 d}\Bigg) \left(\frac{l}{z_H}\right)^{2 d}\Bigg{\}}~,
	      \end{aligned}
	       \end{equation}
where the coefficients $c_k$ are defined in the appendix in (\ref{ck}). These are numerical coefficients depending only on $d$. It is easy to see that there are two expansions in parameters $\frac{\ell}{z_H}$ and $\frac{z_c}{\ell}$ intertwined. These parameters are in fact $T\ell$ and $\frac{\tilde{\lambda}}{\ell}$ which are small in this subsection's limit. Interestingly one can observe that another parameter appears naturally here which is $\frac{z_c}{z_H}$ which makes it three expansions intertwined. We will discuss this point more after the result for the entanglement entropy.

Next step is to expand the relation for area (\ref{Area of entangling region}) similarly to what we did for $z_*$. Keeping terms up to order $2d$ in expansion parameters together and make some simplifications we get
  \begin{equation}
  \begin{aligned}
        A=2 R^{d-1} \left(\frac{L}{z_*}\right)^{d-2}&\bigg{\{}\frac{c_0}{2-d}+\frac{\left(d+1\right)c_1}{4} \left(\frac{z_*}{z_H}\right)^d+\frac{3\left(2 d + 1\right)c_2}{8\left(d + 2\right)} \left(\frac{z_*}{z_H}\right)^{2 d} -\frac{1}{2 d}\left(\frac{z_c}{z_*}\right)^{d}\\
        & -\frac{1}{8 d}\left(\frac{z_*}{z_H}\right)^{d}\left(\frac{z_c}{z_*}\right)^{2 d}
        -\frac{3}{8\left(d + 2\right)}\left(\frac{z_*}{z_H}\right)^{2d}\left(\frac{z_c}{z_*}\right)^{2 d}\bigg{\}}\\
        + 2 R^{d-1} \left(\frac{L}{z_c}\right)^{d-2}
        &\bigg{\{}\frac{1}{d - 2}-\frac{1}{4}\left(\frac{z_c}{z_H}\right)^{d}\bigg{\}}~.
  \end{aligned}
  \end{equation}
 The terms involving $z_*$ in the above relation can be replaced using (\ref{zstar}). After some cumbersome calculation the result for entanglement entropy in low temperature and small deformation limit becomes
\begin{equation}
\begin{aligned}
\label{firstlimit}
  S_{A}=&\frac{2R^{d-1}}{4G_N^{(d+1)}}~\left(\frac{L}{z_c}\right)^{d-2}\Bigg\{\frac{1}{d-2}-\frac{1}{4}\left(\frac{z_c}{z_H}\right)^{d}-\frac{3}{8\left(d+2\right)}\left(\frac{z_c}{z_H}\right)^{2d}+\mathcal{O}\left(\frac{z_c}{z_H}\right)^{3d}\Bigg\}\\
  +&\frac{2R^{d-1}}{4G_N^{(d+1)}}~\left(\frac{L}{\ell}\right)^{d-2}\Bigg\{\left(a_1+ a_5 \left(\frac{z_c}{z_H}\right)^d+a_9 \left(\frac{z_c}{z_H}\right)^{2d}+\mathcal{O}\left(\frac{z_c}{z_H}\right)^{3d}\right)\\
  +&\left(\frac{z_c}{l}\right)^d  \left(a_2+a_6 \left(\frac{z_c}{z_H}\right)^{d}+\mathcal{O}\left(\frac{z_c}{z_H}\right)^{2d}\right)+ \left(\frac{z_c}{\ell}\right)^{2d}\left(a_3+\mathcal{O}\left(\frac{z_c}{z_H}\right)^{d}\right)\\
  +&\left(\frac{\ell}{z_H}\right)^d\left(a_4+a_8 \left(\frac{z_c}{z_H}\right)^d+\mathcal{O}\left(\frac{z_c}{z_H}\right)^{2d} \right)
  +\left(\frac{\ell}{z_H}\right)^{2d} \left(a_7+ \mathcal{O}\left(\frac{z_c}{z_H}\right)^{d} \right)\Bigg\}~,
    \end{aligned}
    \end{equation}
where coefficients $a_{i}$ are introduced in the appendix in (\ref{as}). Similar to $c_k$ they are just numerical factors which depend only on the dimension of field theory spacetime. Looking at this result more closely shows that although the expansion parameters were chosen $z_c/z_*$ and $z_*/z_H$ the final result involves three physical expansion parameters instead of two which are $(\ell/z_H)^d$, $(z_c/\ell)^d$ and $(z_c/z_H)^d$. The last ratio emerges naturally from the calculations which is small in all the limits we consider. In fact $z_H$ puts an upper bound on $z_c$ as the cut-off can not pass the horizon. The ratio $z_c/z_H$ which translates to $\tilde{\lambda} T$ in deformed field theory depends only on field theory parameters, temperature and deformation parameter. It does not relate to the non-local operator we are discussing here and its scale which is the entangling length $\ell$. 

This result can be discussed in different ways. The first line in this result are the corrections due to both finite temperature and finite cut-off to the divergent piece in the limit $z_c\to 0$. The only expansion parameter appearing in the first line and for this set of corrections is $\left(z_c/z_H\right)^d$. It tells us that the corrections to the divergent piece are independent of entangling length $\ell$ and only depend on the information intrinsic to field theory which are deformation parameter $\tilde{\lambda}$ and temperature $T$. Interestingly these corrections all follow area law behaviour for the entanglement entropy but are not divergent.

Another observation is that the set of corrections in the first line are repeated in the last three lines but with different numerical coefficients and different overall factors. In fact for each power of $(z_c/\ell)^d$ and $(\ell/z_H)^d$ the whole expansion in terms of $(z_c/z_H)^d$ appears. Note that the last three lines are corrections which in addition to temperature and deformation parameter are sensitive to how large the entangling length is compared to these parameters. These are in fact the non-local effects appearing in the entanglement entropy.

 To check whether our result produces the already known results in the literature we first take the limit where the cut-off goes to zero that is $z_c$ or $\tilde{\lambda} \to 0$. The entropy relation reduces to
\begin{equation}
\label{zerocutoff}
  S_{A}|_{z_c\to 0}=\frac{2R^{d-1}}{4G_N^{(d+1)}}~\Bigg{\{}\left(\frac{L}{z_c}\right)^{d-2}\frac{1}{d-2}
  +\left(\frac{L}{\ell}\right)^{d-2}\left(a_1+a_4 \left(\frac{\ell}{z_H}\right)^d
  +a_7\left(\frac{\ell}{z_H}\right)^{2d}\right) \Bigg\}~.
    \end{equation}
    This is exactly what we expected for the field theory which lives on the boundary as in \cite{Fischler:2012ca}. The first term is the divergent piece which appears in the limit where $T\to 0$ or $z_H \to \infty$. Note that as we expect the divergent piece is proportional to the area of the subsystem. Replacing our background with $AdS_5$ black brane solution the entanglement entropy in (\ref{zerocutoff}) reduces to 
  \begin{equation}
  \label{zeroT}
 S_{A}=\frac{R^3}{4G_N^{(5)}}~\Bigg\{\left(\frac{L}{z_c}\right)^2+ \left(\frac{L}{\ell}\right)^{2}  \Bigg[-0.32+ 0.56 \left(\frac{l}{z_H}\right)^4\Bigg]\Bigg\}~,
 \end{equation}
which is exactly what we must get for this background \cite{Fischler:2012ca}. 

If we compare relation (\ref{zerocutoff}) with the original result (\ref{firstlimit}) we see that in the field theory at finite temperature introducing deformation parameter modifies the entanglement entropy in two ways: set of corrections in terms of $(\tilde{\lambda} T)^d$ and set of corrections in terms of $(\tilde{\lambda}/\ell)^d$. Such structure repeats itself when we add temperature to a deformed theory. The only difference is that the corrections are in terms of $(\tilde{\lambda} T)^d$ and $(\ell T)^d$. To see this more explicitly we take the limit where $z_H \to \infty$ or $T\to 0$. This gives us the entanglement entropy in a $T\bar{T}$ deformed field theory at zero temperature. The relation (\ref{firstlimit}) reduces to 
\begin{equation}
  S_{A}|_{T\to 0}=\frac{2R^{d-1}}{4G_N^{(d+1)}}~\Bigg\{\left(\frac{L}{z_c}\right)^{d-2}\frac{1}{d-2}
  + \left(\frac{L}{\ell}\right)^{d-2}\left(a_1
  +a_2 \left(\frac{z_c}{l}\right)^d+a_3 \left(\frac{z_c}{\ell}\right)^{2d}\right)\Bigg\}~.
    \end{equation}
This result, up to order $(z_c/\ell)^d$, matches what has been obtained in \cite{Khoeini-Moghaddam:2020ymm} if their background is reduced to $AdS_{d+1}$ with non-zero cut-off. It can be easily seen that the corrections to the entanglement entropy at non-zero cut-off are again proportional to the area and appear similarly to the corrections due to the non-zero temperature at (\ref{zerocutoff}). For $AdS_5$ black brane solution this relation reduces to
  \begin{equation}
S_{A}=\frac{R^3}{4G_N^{(5)}}~\Bigg\{\left(\frac{L}{z_c}\right)^2+\left(\frac{L}{l}\right)^{2} \left(-0.32+0.24\left(\frac{z_c}{l}\right)^4 \right)\Bigg\}~.
    \end{equation}
It somehow tells us that the entanglement entropy in field theories with holographic duals react to the non-zero deformation parameter similarly to the temperature but regarding the numerical coefficients the effect of temperature on entanglement entropy is slightly bigger than cut-off.

Note that these observations are true for the limit where $T\ell \ll 1$ and $\tilde{\lambda}/\ell \ll 1$. In the next subsections we will see whether such results can be derived in the other limits too.

\subsection{Low temperature and large deformation limit; \textbf{$z_c \simeq z_* \ll z_H$}}
\label{low large}

This limit tells us that the entangling surface probes the region near the cut-off which is far from the horizon. In field theory language the parameters in this limit go as $T\ell \ll 1$ and $\tilde{\lambda}/\ell \gg 1$. It suggests that the entangling surface sees the effect of the non-zero cut-off more than the effect of the horizon. In other words the entanglement entropy is more sensitive to the non-zero deformation parameter than to the non-zero temperature in this limit. To clarify more on these points let us go back to the relations for $\ell$ (\ref{ell2}) and the area (\ref{area2}) and  and see how they behave in the limit where $z_c/z_* \simeq 1$ and $z_*/z_H \ll 1$.
	
The first obstacle we face is that the argument in the hypergeometric functions is not small any more and one needs to check whether the series in the area and $\ell$ are convergent. Hypergeometric functions of the form $_2F_1(a,b,c;z)$ are absolutely convergent near $z=1$ if $c-a-b$ is not an integer and $Re(c-a-b)>0$. In such condition the hypergeometric function can be written as a sum of two other hypergeometric functions with arguments $1-z$ which are solutions to the Euler's hypergeometric differential equation near $z=1$. Such identity helps us to write the hypergeometric functions as series in the small expansion parameter $1-z$. The exact identity is given in (\ref{hypid}). Since the combination $c-a-b$ in the hypergeometric functions we have in (\ref{ell2}) and (\ref{area2}) satisfies the above conditions the relations for area and $\ell$ are convergent. Therefore for the limit $z_c/z_* \simeq 1$ we can use this identity and write (\ref{ell2}) and (\ref{area2}) as	
	\begin{equation}\label{ell4}
  \begin{aligned}
  \frac{l}{2} &= z_* \sum_{k=0}^{\infty} \frac{1}{1+kd} ~\frac{\Gamma(k+\frac{1}{2})~\Gamma\left(\frac{d(k+1)}{2(d-1)}\right)}{\Gamma(k+1)~\Gamma\left(\frac{1+kd}{2(d-1)}\right)} \left(\frac{z_*}{z_H}\right)^{kd}\\
	       &- z_* \sum_{k=0}^{\infty} \frac{\Gamma(k+\frac{1}{2})}{\sqrt{\pi}\Gamma(k+1)}~\left(\frac{z_*}{z_H}\right)^{kd}~\left(\frac{z_c}{z_*}\right)^{d(k+1)}\times \Biggl{\{}~\frac{\sqrt{\pi}}{1+kd}~\frac{\Gamma\left(\frac{d(k+1)}{2(d-1)}\right)}{\Gamma\left(\frac{1+kd}{2(d-1)}\right)}
~_2F_1\left(\frac{1}{2},\frac{d(k+1)}{2(d-1)};\frac{1}{2},1-\left(\frac{z_c}{z_*}\right)^{2(d-1)}\right)\\
          &-\frac{2}{2(d-1)}~\sqrt{1-\left(\frac{z_c}{z_*}\right)^{2(d-1)}}~_2F_1\left(\frac{d(k+2)-1}{2(d-1)},1;\frac{3}{2},1-\left(\frac{z_c}{z_*}\right)^{2(d-1)}\right)~\Biggr{\}}~,
  \end{aligned}
\end{equation}
and
\begin{equation}\label{area4}
  \begin{aligned}
  \mathcal{A} &= 2R^{d-1}\left(\frac{L}{z_*}\right)^{d-2} \sum_{k=0}^{\infty} \frac{1}{2(d-1)} ~\frac{\Gamma(k+\frac{1}{2})~\Gamma\left(\frac{d(k-1)+2}{2(d-1)}\right)}{\Gamma(k+1)~\Gamma\left(\frac{1+kd}{2(d-1)}\right)} \left(\frac{z_*}{z_H}\right)^{kd}\\
	       &- 2R^{d-1}\left(\frac{L}{z_*}\right)^{d-2} \sum_{k=0}^{\infty} \frac{\Gamma(k+\frac{1}{2})}{\sqrt{\pi}\Gamma(k+1)}\frac{1}{2+d(k-1)}~\left(\frac{z_*}{z_H}\right)^{kd}~\left(\frac{z_c}{z_*}\right)^{d(k-1)+2}\times \\
& \Biggl{\{}~\frac{\sqrt{\pi}\Gamma\left(\frac{d(k+1)}{2(d-1)}\right)}{\Gamma\left(\frac{1+kd}{2(d-1)}\right)}
~_2F_1\left(\frac{1}{2},\frac{d(k-1)+2}{2(d-1)};\frac{1}{2},1-\left(\frac{z_c}{z_*}\right)^{2(d-1)}\right)\\
          &-2\frac{\Gamma\left(\frac{d(k+1)}{2(d-1)}\right)}{\Gamma\left(\frac{d(k-1)+2}{2(d-1)}\right)}~\sqrt{1-\left(\frac{z_c}{z_*}\right)^{2(d-1)}}~
          _2F_1\left(\frac{kd+1}{2(d-1)},1;\frac{3}{2},1-\left(\frac{z_c}{z_*}\right)^{2(d-1)}\right)~\Biggr{\}}~.
  \end{aligned}
\end{equation}
Interestingly another hypergeometric function identity can help us to simplify these relations tremendously. For hypergeometric function $_2F_1(a,b,c;z)$ if $a=c$ the hypergeometric function can be written as $(1-z)^{c-a-b}$. As one can easily see the hypergeometric functions in the second line of $\ell$ (\ref{ell4}) and the third line of the area (\ref{area4}) have such form and therefore they are cancelled by the terms in the first line. Therefore the relations for area and entangling length reduces to
	\begin{equation}\label{ell42}
  \begin{aligned}
  \frac{l}{2} &= \frac{z_*}{(d-1)}\sum_{k=0}^{\infty} \frac{\Gamma(k+\frac{1}{2})}{\sqrt{\pi}\Gamma(k+1)}~\left(\frac{z_*}{z_H}\right)^{kd}~\left(\frac{z_c}{z_*}\right)^{d(k+1)}\\
          &\times \sqrt{1-\left(\frac{z_c}{z_*}\right)^{2(d-1)}}~_2F_1\left(\frac{d(k+2)-1}{2(d-1)},1;\frac{3}{2},1-\left(\frac{z_c}{z_*}\right)^{2(d-1)}\right)~,
  \end{aligned}
\end{equation}
and
\begin{equation}\label{area42}
  \begin{aligned}
  \mathcal{A} &=  2R^{d-1}\left(\frac{L}{z_*}\right)^{d-2} \sum_{k=0}^{\infty} \frac{\Gamma(k+\frac{1}{2})}{\sqrt{\pi}\Gamma(k+1)}\frac{1}{d-1}~\left(\frac{z_*}{z_H}\right)^{kd}~\left(\frac{z_c}{z_*}\right)^{d(k-1)+2} \\
& \times \sqrt{1-\left(\frac{z_c}{z_*}\right)^{2(d-1)}}~
          _2F_1\left(\frac{kd+1}{2(d-1)},1;\frac{3}{2},1-\left(\frac{z_c}{z_*}\right)^{2(d-1)}\right)~.
  \end{aligned}
\end{equation}
To obtain the area in this limit we should follow the same path as the previous subsection. We have to obtain $z_*$ in terms of $\ell$ and substitute it in the relation for the area. But since we have $z_* \simeq z_c$ here we can define a new small parameter and expand hypergeometric function in (\ref{ell42}) in terms of it using (\ref{hypergeometric function}). We can write $z_*=z_c (1+\epsilon)$ where $\epsilon \equiv \frac{z_*-z_c}{z_c}$ and goes to zero as $z_*$ approaches $z_c$.
Therefore the argument in the hypergeometric function up to second order in $\epsilon$ for $\epsilon \ll 1$ reduces to
\begin{equation}\label{epsilon_expansion}
  1-\left(\frac{z_c}{z_*}\right)^{2(d-1)} \simeq  2(d-1)\epsilon\left[1 -\frac{1}{2}(2d-1)\epsilon+\dots \right]~.
\end{equation}
Writing the fraction $z_c/z_*$ in terms of small parameter $\epsilon$ helps us expand the hypergeometric functions in (\ref{ell42}) and (\ref{area42}) using (\ref{hypergeometric function}). Therefore instead of writing $z_*$ in terms of $\ell$ we obtain $\epsilon$ and substitute it in the area to derive the entanglement entropy. An interesting observation is that in the relations for the area and $\ell$ all the integer powers of $\epsilon$ are cancelled and we are only left with half-integer powers of it. Note that we keep terms up to order $5/2$ in $\epsilon$ in the relation for $\ell$. This choice produces the expansions in the final results up to the total order $2d$ which we demand. If we solve $\ell$ for $\epsilon$ perturbatively we get
\begin{equation}
\label{epsilonresult}
\sqrt{\epsilon} = \frac{\ell}{2 z_c} \sqrt{\frac{(d-1)}{2}} \mathcal{S}_1^{-1} \Bigg{\{}1-\left(\frac{\ell}{2 z_c}\right)^2 \frac{(d-1)}{2}~\mathcal{S}_1^{-3}\mathcal{S}_2 +\left(\frac{\ell}{2 z_c}\right)^4 \frac{(d-1)^2}{4}~\mathcal{S}_1^{-6} \left(3 \mathcal{S}_2^2-\mathcal{S}_1\mathcal{S}_3\right)\Bigg\}~,
\end{equation}
where
\begin{equation}
\begin{aligned}
\mathcal{S}_1 =& \sum^\infty_{n=0}  \frac{\Gamma\left(n+\frac{1}{2}\right)}{\sqrt{\pi}\Gamma\left(n+1\right)} \left(\frac{z_c}{z_H}\right)^{n d}~,\\
\mathcal{S}_2 =&\sum^\infty_{k=0} \frac{\Gamma\left(k+\frac{1}{2}\right)}{\sqrt{\pi} \Gamma\left(k+1\right)} \frac{-2 d+7 +8 k d}{12} \left(\frac{z_c}{z_H}\right)^{k d}~,\\
\mathcal{S}_3 =&\sum^\infty_{k=0} \frac{\Gamma\left(k+\frac{1}{2}\right)}{\sqrt{\pi} \Gamma\left(k+1\right)} \frac{-71+4 d (5+12 k)+4 d^2 (1-8 k+32 k^2)}{480} \left(\frac{z_c}{z_H}\right)^{k d}~.
\end{aligned}
\end{equation}
Note that the calculation itself tells us that $\epsilon$ is an expansion in terms of two expansion parameters $\ell/ z_c$ and $z_c/z_H$. We just remind the reader that we are working in the limit where $\ell/\tilde{\lambda}\ll 1$ or in other words $\ell /z_c \ll 1$ and $z_c/z_H\ll 1$. As one can see, the extra expansion parameter $\ell / z_H$ does not appear in the result here. This is somehow expected since we are probing the region near the cut-off and the temperature affects the result only through the ratio $z_c/z_H$ or $\tilde{\lambda} T$. Replacing $\epsilon$ in the relation for $z_*$ we have
\begin{equation}
\begin{aligned}
  z_*=z_c~\Bigg{\{}1+&~\frac{d-1}{8}\left(1-\left(\frac{z_c}{z_H}\right)^d~\right)~\left(\frac{l}{z_c}\right)^2\\
  +& \frac{(d-1)^2~(2d-7)}{384}~\left(1-2~ \frac{4d-7}{2d-7}~\left(\frac{z_c}{z_H}\right)^d+~ \frac{6d-7}{2d-7}~\left(\frac{z_c}{z_H}\right)^{2d}~\right)\left(\frac{l}{z_c}\right)^4\\
  +& \frac{(d-1)^3~(241-130 d+16 d^2)}{46080} \left(\frac{l}{z_c}\right)^6\\
  \times&\left(1-\frac{723-626 d+136 d^2}{241-130 d+16 d^2}~\left(\frac{z_c}{z_H}\right)^d+\frac{723-862 d+258 d^2}{241-130 d+16 d^2}~\left(\frac{z_c}{z_H}\right)^{2d}-\frac{241-366 d+138 d^2}{241-130 d+16 d^2}~\left(\frac{z_c}{z_H}\right)^{3d}\right)\Bigg\}~.
\end{aligned}
\end{equation}
 An important point that we should mention is that $\epsilon$ is a positive parameter so that $z_*$ reaches $z_c$ from above and never crosses the cut-off. In addition to not having the expansion parameter $\ell/z_H$, another major difference between this result and the result for $z_*$ in the previous subsection (\ref{zstar}) is the order of the expansion parameters. In fact we see the parameter $z_c/z_H$ appears in orders of power $d$ as one could expect compared to the previous subsection. But the powers of the $\ell/z_c$ parameter are integer powers of 2. We will elaborate on this point more later when we obtain the entanglement entropy.

 Now we are ready to calculate the area using the results for $z_*$ and $\epsilon$. Similar to $\ell$ all integer powers of $\epsilon$ are cancelled in the relation for area and we are left with half-integer powers of it. The result for the area after some simplification is
\begin{equation}
\begin{aligned}
\label{area43}
   \mathcal{A}= &2R^{d-1}\left(\frac{L}{z_c}\right)^{d-2}\sqrt{\frac{2}{(d-1)}} \sum_{k=0}^{\infty}\frac{\Gamma(k+\frac{1}{2})}{\sqrt{\pi}\Gamma(k+1)}\left(\frac{z_c}{z_H}\right)^{kd}\\
  &\times\sqrt{\epsilon} \left[1+\frac{\epsilon}{12} \left(2d\left(4k-3\right)+11\right) +\frac{\epsilon^2}{480} \left(97+ 4 d \left(-61+60 k\right) + 4 d^2 \left(25-56 k+32 k^2\right)\right)+\mathcal{O}(\epsilon^3)\right]~,
\end{aligned}
\end{equation}
where we have kept terms up to order $\epsilon^{5/2}$. Following the same procedure, we can again replace $\epsilon$ in the area with the relation (\ref{epsilonresult}) and the final result for entanglement entropy becomes
\begin{equation}
\label{secondlimit}
\begin{aligned}
S_{A} = \frac{R^{d-1}}{4G_N^{(d+1)}}\left(\frac{L}{z_c}\right)^{d-2}\bigg\{&\left(\frac{\ell}{z_c}\right)-\frac{(d-1)^2}{24}
\left(1-\left(\frac{z_c}{z_H}\right)^d\right)~\left(\frac{\ell}{z_c}\right)^3\\
+&\frac{(d-1)^3}{1920}(d+7)
\left(1+ 2 \frac{d-7}{d+7}\left(\frac{z_c}{z_H}\right)^d-\frac{3d-7}{d+7}\left(\frac{z_c}{z_H}\right)^{2d}\right)~\left(\frac{\ell}{z_c}\right)^5+\mathcal{O}~\left(\frac{\ell}{z_c}\right)^7\bigg\}~.
\end{aligned}
\end{equation}
Due to the half-integer powers of $\epsilon$ in the expansion (\ref{area43}) we just have odd powers of $\ell/z_c$ without any dimension dependence in its powers. Therefore the expansion in $\ell/z_c$ or $\ell/\tilde{\lambda}$ is universal and appears in any deformed field theory with holographic dual. Only the coefficients depend on dimension of the spacetime.  As mentioned previously this is another key distinction between this and the last limit for the entanglement entropy. Note that all the terms involve the overall coefficient $\left(L/z_c\right)^{d-2}$ which is the area law divergent piece at $z_c\to 0$. Therefore in the large deformation and small temperature regime the entanglement entropy shows volume law behaviour, similar to high temperature limit where the thermal entropy dominates. This tells us that deformation and temperature affect the entanglement entropy in a similar way.

Let's take the limit where $z_c \to 0$ or zero cut-off limit. Since we are working in the limit where $\ell/\tilde{\lambda} \ll 1$, at $z_c \to 0$, we have $\ell \to 0$ in a way that the fraction $\ell / z_c$ remain finite and small. Therefore the only divergent piece is overall factor $\left(L/z_c\right)^{d-2}$ in the first line. 

In the zero temperature limit where $z_H \to \infty$ the ratio $z_c/z_H$ goes to zero and we are left with a power series in $\ell/z_c$ with odd powers. This matches the result in \cite{Khoeini-Moghaddam:2020ymm} up to order $(\ell/z_c)^3$ if their background is reduced to $AdS_{d+1}$ with non-zero cut-off.

Another interesting observation is that the terms involving temperature correction as $\left(z_c/z_H\right)^d$ appear at order $\left(\ell/z_c\right)^3$ and  $\left(z_c/z_H\right)^{2d}$ appears at order  $\left(\ell/z_c\right)^5$. It shows that higher order corrections in temperature appear at higher order corrections of the cut-off. It somehow says that as $z_*$ gets farther away from the cut-off it sees more of the horizon and higher order corrections due to temperature show up. This coincides with what we expect from physics perspective.

\subsection{High temperature and small deformation limit; \textbf{$z_c \ll z_* \simeq z_H$}}
\label{high small}
In the deformed field theory language this limit corresponds to $T\ell \gg 1$ and $\tilde{\lambda}/\ell \ll 1$. In fact if we look at the relations for $\ell$ and the area in (\ref{ell2}) and (\ref{area2}) we will see that the hypergeometric functions have small arguments and can be expanded while the infinite series in $z_*/z_H$ should be examined whether they are convergent since $z_*/z_H \to 1$. To check the convergence of the series we take their large $k$ limit and see how they behave. The large $k$ limit of the series in (\ref{ell2}) and (\ref{area2}) gives $\frac{1}{k}\left(\frac{z_*}{z_H}\right)^{kd}$. Therefore we have
 \begin{equation}
\sum_{k=1}^\infty \frac{1}{k}\left(\frac{z_*}{z_H}\right)^{kd} =  - \text{ln} \left(1-\left(\frac{z_*}{z_H}\right)^d\right)~,
\end{equation}
which is divergent for $\left(z_*/z_H\right)^d \to 1$. The promising point is that this divergence is the same in both area and $\ell$ and therefore we can write area in terms of $\ell$. This helps us isolate the divergence in area using the isolation of divergence in $\ell$. After some tedious calculation we write the area in terms of $\ell$ in this limit as
\begin{equation}
  \begin{aligned}
  \label{Athirdlimit}
  \mathcal{A}=&2R^{d-1}\left(\frac{L}{z_*}\right)^{d-2}~\Biggl{\{} \frac{\ell}{2z_*}+
  \sum_{k=0}^{\infty}\frac{d-1}{(kd+1)(d(k-1)+2)}\frac{\Gamma\left(k+\frac{1}{2}\right)\Gamma\left(\frac{d(k+1)}{2(d-1)}\right)}{\Gamma\left(k+1\right)\Gamma\left(\frac{kd+1}{2(d-1)}\right)}\left(\frac{z_*}{z_H}\right)^{kd}\\ +&\sum_{k=0}^{\infty}\frac{1}{\sqrt{\pi}d(k+1)}\frac{\Gamma(k+\frac{1}{2})}{\Gamma(k+1)}\left(\frac{z_*}{z_H}\right)^{kd}\left(\frac{z_c}{z_*}\right)^{d(k+1)}~
  _2F_1\left(\frac{1}{2},\frac{d(k+1)}{2(d-1)};\frac{d(k+3)-2}{2(d-1)},\left(\frac{z_c}{z_*}\right)^{2(d-1)}\right)\\
  -&\sum_{k=0}^{\infty}\frac{1}{\sqrt{\pi}(d(k-1)+2)}\frac{\Gamma(k+\frac{1}{2})}{\Gamma(k+1)}\left(\frac{z_*}{z_H}\right)^{kd}\left(\frac{z_c}{z_*}\right)^{d(k-1)+2}~ _2F_1\left(\frac{1}{2},\frac{d(k-1)+2}{2(d-1)};\frac{d(k+1)}{2(d-1)},\left(\frac{z_c}{z_*}\right)^{2(d-1)}\right)\Biggr{\}}~.
  \end{aligned}
\end{equation}
A a result the first line in area (\ref{area2}) which is an infinite series in terms of $\left(\frac{z_*}{z_H}\right)^d$ parameter becomes convergent since for large $k$ it behaves as  $\frac{1}{k^2}\left(\frac{z_*}{z_H}\right)^{kd}$ and
\begin{equation}
\sum_{k=1}^\infty \frac{1}{k^2}\left(\frac{z_*}{z_H}\right)^{kd}=\text{polylog}\left(2,\left(\frac{z_*}{z_H}\right)^{d}\right)  \equiv Li_2\left(\left(\frac{z_*}{z_H}\right)^d\right)~,
\end{equation}
 is convergent even for $z_*=z_H$. More details on polylog function can be found in appendix (\ref{eq18b}). Now all the divergence for infinite series rests in $\ell$ and to have a finite area we just need to regularize $\ell$. Going back to the relation for $\ell$ in (\ref{ell2}) we can regularize it by adding and subtracting the corresponding divergent series. Therefore relation (\ref{ell2}) becomes
\begin{equation}
  \begin{aligned}\label{isolating_l}
    \ell=&2c_0~z_* + z_* \sum_{k=1}^{\infty}\left[\frac{\Gamma\left(k+\frac{1}{2}\right)\Gamma\left(\frac{d(k+1)}{2(d-1)}\right)}{\Gamma\left(k+1\right)\Gamma\left(\frac{kd+1}{2(d-1)}\right)}\frac{2}{kd+1}-\sqrt{\frac{2}{d(d-1)}}\frac{1}{k}\right]
    \left(\frac{z_*}{z_H}\right)^{kd}-z_* \sqrt{\frac{2}{d(d-1)}}\ln \left[1-\left(\frac{z_*}{z_H}\right)^d\right] \\
    - & 2z_* \sum_{k=0}^{\infty}\frac{1}{\sqrt{\pi}~d(k+1)}\frac{\Gamma(k+\frac{1}{2})}{\Gamma(k+1)}\left(\frac{z_*}{z_H}\right)^{kd}\left(\frac{z_c}{z_*}\right)^{d(k+1)}~ _2F_1\left(\frac{1}{2},\frac{d(k+1)}{2(d-1)};\frac{d(k+3)-2}{2(d-1)},\left(\frac{z_c}{z_*}\right)^{2(d-1)}\right)~.
  \end{aligned}
\end{equation}
Since in this limit we have $\frac{z_*}{z_H}\simeq 1$ we can write $z_*=z_H(1-\epsilon)$ where $\epsilon=\frac{z_H-z_*}{z_H}$ is small and should be positive. Positivity of $\epsilon$ means that the extremal surface does not penetrate into the horizon which is what we expect. To see this we replace the relation for $z_*$ in terms of $\epsilon$ into $\ell$ (\ref{isolating_l}) and solve it for $\epsilon$. The result is
\begin{equation}\label{epsilon}
\begin{aligned}
  \epsilon=\epsilon_d~\exp \Bigg[&-\sqrt{\frac{d(d-1)}{2}}\Biggl{\{}\frac{l}{z_H}\\
  &+2\sum_{k=0}^{\infty}\frac{1}{\sqrt{\pi}~d(k+1)}\frac{\Gamma(k+\frac{1}{2})}{\Gamma(k+1)}\left(\frac{z_c}{z_H}\right)^{d(k+1)}~ _2F_1\left(\frac{1}{2},\frac{d(k+1)}{2(d-1)};\frac{d(k+3)-2}{2(d-1)},\left(\frac{z_c}{z_H}\right)^{2(d-1)}\right)\Biggr{\}} \Bigg]~,
\end{aligned}
\end{equation}
where
\begin{equation}
  \epsilon_d=\frac{1}{d}\exp \left[\sqrt{\frac{d(d-1)}{2}}\Biggl{\{} 2c_0 +\sum_{k=1}^{\infty}\left(\frac{\Gamma\left(k+\frac{1}{2}\right)\Gamma\left(\frac{d(k+1)}{2(d-1)}\right)}{\Gamma\left(k+1\right)\Gamma\left(\frac{kd+1}{2(d-1)}\right)}\frac{2}{kd+1}-\sqrt{\frac{2}{d(d-1)}}\frac{1}{k}\right)\Biggr{\}}\right]~.
\end{equation}
$c_0$ is the numerical coefficient defined in (\ref{ck}). Now that we have expansion of $\epsilon$ in terms of integer powers of $z_c/z_H$ and $\ell/z_H$ we can write $z_*$ in terms of them using $z_*=z_H(1-\epsilon)$. Since it is a straightforward calculation we did not bring the final result for $z_*$ here. Important point here is that as we expected $\epsilon$ is positive and the extremal surface never crosses the horizon.
%
%

So far we have obtained $\epsilon$ and therefore $z_*$ in terms of our physical expansion parameters. We have also isolated the divergent series in $\ell$ and made it convergent. We can now check the area and write it in terms of orders of small parameter $\epsilon$. To do so, we write the area as sum of two parts $\mathcal{A}_1$ and $\mathcal{A}_2$ where $\mathcal{A}_1$ does not include the hypergeometric terms and $\mathcal{A}_2$ only includes the hypergeometric ones. The leading order term for $\mathcal{A}_1$ is obtained by replacing $z_*$ with $z_H$. To obtain terms of order $\epsilon$, similarly to $\ell$, we again write the series in $\mathcal{A}_1$ for large $k$ and add and subtract the corresponding series like what we did in regularizing $\ell$. Therefore we have
\begin{equation}\label{isolating_A_1}
\begin{aligned}
  \mathcal{A}_1 & = 2R^{d-1}\left(\frac{L}{z_*}\right)^{d-2}\Biggl{\{}\left(\frac{l}{2z_*}\right)-\frac{d-1}{d-2}c_0\\
      & + \sum_{k=1}^{\infty}\left[\frac{\Gamma\left(k+\frac{1}{2}\right)\Gamma\left(\frac{d(k+1)}{2(d-1)}\right)}{\Gamma\left(k+1\right)\Gamma\left(\frac{kd+1}{2(d-1)}\right)}\frac{d-1}{(kd+1)(d(k-1)+2)}-\sqrt{\frac{d-1}{2d}}\frac{1}{dk^2}\right]\left(\frac{z_*}{z_H}\right)^{kd}\\
      & +\frac{1}{d}\sqrt{\frac{d-1}{2d}}Li_2 \left[\left(\frac{z_*}{z_H}\right)^d\right]\Biggr{\}}~.
\end{aligned}
\end{equation}
Note that, in contrast to $\ell$, the large k limit of the infinite series in $\mathcal{A}_1$ produces convergent series even in the limit $z_* \simeq z_c$ as $\sum_{k=0}^\infty \frac{1}{k^2} = \frac{\pi^2}{6}$. It is in fact the polylogarithm function of order 2 and argument $z_*/z_H$, last line of the above equation, defined in (\ref{polylog}).  Now we can replace $z_*$ with $z_H(1-\epsilon)$ in this equation and use the fact that
\begin{equation}
\lim_{\epsilon \to 0} Li_2(\frac{z_*}{z_H}) \simeq \frac{\pi^2}{6}+d\left(\text{ln} (d \epsilon) -1\right)\epsilon~,
\end{equation}
up to order $\epsilon$.
Therefore the relation for $A_1$ reduces to
\begin{equation}
  \begin{aligned}
    \mathcal{A}_1  &= 2 R^{d-1} \left(\frac{L}{z_H}\right)^{d-2}\Biggl{\{}\frac{\ell}{2 z_H}- \sqrt{\frac{d-1}{2 d}}~\epsilon +\hat{\mathcal{S}}\\
    &- (d-1) \epsilon \sum_{k=0}^{\infty}\frac{1}{\sqrt{\pi}d(k+1)}\frac{\Gamma(k+\frac{1}{2})}{\Gamma(k+1)}\left(\frac{z_c}{z_H}\right)^{d(k+1)}~ _2F_1\left(\frac{1}{2},\frac{d(k+1)}{2(d-1)};\frac{d(k+3)-2}{2(d-1)},\left(\frac{z_c}{z_H}\right)^{2(d-1)}\right)\Biggr{\}}~,
  \end{aligned}
\end{equation}
where we have kept terms up to order $\epsilon$.  We have denoted all numerical terms which depend only on $d$ by $\hat{\mathcal{S}}$ that is
\begin{equation}
\hat{\mathcal{S}} = - \frac{d-1}{d-2}c_0
     + \sum_{k=1}^{\infty}\frac{\Gamma\left(k+\frac{1}{2}\right)\Gamma\left(\frac{d(k+1)}{2(d-1)}\right)}{\Gamma\left(k+1\right)\Gamma\left(\frac{kd+1}{2(d-1)}\right)} \frac{(d-1)}{(1+ k d) \left(d\left(k-1\right)+2\right)}~.
\end{equation}
As mentioned previously we called the part of area that includes only the hypergeometric terms, the last two lines of (\ref{thirdlimit}), $\mathcal{A}_2$.  We can again replace $z_*$ with its definition in terms of $\epsilon$ and therefore $\mathcal{A}_2$ reduces to
\begin{equation}
  \begin{aligned}
A_2&=2R^{d-1}\left(\frac{L}{z_H}\right)^{d-2}~\Biggl{\{} \sum_{k=0}^{\infty}\frac{1}{\sqrt{\pi}d(k+1)}\frac{\Gamma(k+\frac{1}{2})}{\Gamma(k+1)}\left(\frac{z_c}{z_H}\right)^{d(k+1)}~ _2F_1\left(\frac{1}{2},\frac{d(k+1)}{2(d-1)};\frac{d(k+3)-2}{2(d-1)},\left(\frac{z_c}{z_H}\right)^{2(d-1)}\right)\\
&-\sum_{k=0}^{\infty}\frac{1}{\sqrt{\pi}\left(2+d\left(k-1\right)\right)}\frac{\Gamma(k+\frac{1}{2})}{\Gamma(k+1)}\left(\frac{z_c}{z_H}\right)^{2+d(k-1)}~ _2F_1\left(\frac{1}{2},\frac{2+d(k-1)}{2(d-1)};\frac{d(k+1)}{2(d-1)},\left(\frac{z_c}{z_H}\right)^{2(d-1)}\right)\Biggr{\}}\\
&+2R^{d-1}\left(\frac{L}{z_H}\right)^{d-2}(d-1)\epsilon~\Biggl{\{} \sum_{k=0}^{\infty}\frac{1}{\sqrt{\pi}d(k+1)}\frac{\Gamma(k+\frac{1}{2})}{\Gamma(k+1)}\left(\frac{z_c}{z_H}\right)^{d(k+1)}~ _2F_1\left(\frac{1}{2},\frac{d(k+1)}{2(d-1)};\frac{d(k+3)-2}{2(d-1)},\left(\frac{z_c}{z_H}\right)^{2(d-1)}\right)\\
&+\sum_{k=0}^{\infty}\frac{2}{\sqrt{\pi}\left(d\left(k+3\right)-2\right)}\frac{\Gamma(k+\frac{1}{2})}{\Gamma(k+1)}\left(\frac{z_c}{z_H}\right)^{d(k+3)-2}~ _2F_1\left(\frac{3}{2},\frac{d(k+3)-2}{2(d-1)};\frac{d(k+5)-4}{2(d-1)},\left(\frac{z_c}{z_H}\right)^{2(d-1)}\right)\Biggr{\}}~.
 \end{aligned}
\end{equation}
Now we can add the two parts $A_1$ and $A_2$ to write the full form of the area in this limit. Since all the infinite series in this result for area are convergent we can expand the hypergeometric functions and make the relation more comprehensible. Note that the argument in the hypergeometric functions is $z_c/z_H$ which is small in this limit. It corresponds to $T \tilde{\lambda} \ll 1$ in the deformed field theory side.

Final result for the entanglement entropy in this limit is
\begin{equation}
\begin{aligned}
\label{thirdlimit}
 S_A &=\frac{2R^{d-1}}{4G_N^{(d+1)}}\left(\frac{L}{z_H}\right)^{d-2}\Biggl{\{}\frac{l}{2z_H}+\hat{\mathcal{S}}-\sqrt{\frac{d-1}{2d}}~\epsilon
  +\frac{1}{2d}\left(\frac{z_c}{z_H}\right)^d
 +\frac{1}{8d}\left(\frac{z_c}{z_H}\right)^{2d}\Biggr{\}}\\
+&\frac{2R^{d-1}}{4G_N^{(d+1)}}\left(\frac{L}{z_c}\right)^{d-2}\Biggl{\{}\frac{1}{d-2}-\frac{1}{4}\left(\frac{z_c}{z_H}\right)^d-\frac{3}{8(d+2)}\left(\frac{z_c}{z_H}\right)^{2 d} \Biggr{\}}~.
\end{aligned}
\end{equation}
Interestingly the terms proportional to $\epsilon$ in $A_2$ get canceled and the only term with $\epsilon$ comes from $A_1$.
Note that in this result two of the expansion parameters of the bulk gravity or deformed field theory appear; $z_c/z_H$ and $\ell/z_H$ or $\tilde{\lambda} T$ and $\ell T$. There are also the expansion parameters in $\epsilon$ as in (\ref{epsilon}). Again the limit itself tells us what combinations of the parameters of the theory should emerge.  In fact in this limit we are probing the region near the horizon and therefore the expansion parameter $z_c/\ell$ does not appear. Comparing this result with the entropy derived in the small deformation and low temperature limit, relation (\ref{firstlimit}), we observe that the terms proportional to $(L/z_c)^{d-2}$ are the same. In fact these terms are the correction terms to the divergent piece in undeformed field theory.  Therefore we can conclude that temperature affects the divergent piece only through the combination $(z_c/z_H)^d$. 

Another observation is that, in the high temperature limit, the first term in the entanglement, relation (\ref{thirdlimit}), behaves as volume of the entangling region which is the behaviour of the thermal entropy. But as we see the rest of the terms which are corrected effect of the temperature by deformation parameter behave as area law. We can compare this with the entropy in the low temperature and large deformation limit, relation (\ref{secondlimit}), where we observed volume law behaviour in the leading correction term to the divergent piece in undeformed theory. This tells us that the leading correction term due to $T$ in the high temperature limit and the leading correction term due to $\tilde{\lambda}$ in the large deformation limit behave similarly.

To check whether our result satisfies the known results in the literature, we send the cut-off in the bulk to the boundary of AdS, $z_c \to 0$, which means we send the deformation parameter in field theory to zero. We obtain 
\begin{equation}
S_A =\frac{2R^{d-1}}{4G_N^{(d+1)}}\left(\frac{L}{z_H}\right)^{d-2}\Biggl{\{}\frac{l}{2z_H}+\hat{\mathcal{S}}-\sqrt{\frac{d-1}{2d}}~\epsilon\Biggr{\}}
  +\frac{2R^{d-1}}{4G_N^{(d+1)}} \frac{1}{d-2} \left(\frac{L}{z_c}\right)^{d-2} ~,
  \end{equation}
which is the result of black brane entropy at high temperature limit \cite{Fischler:2012ca}. The last term is the divergent piece which behaves as the area law. In this relation $\epsilon$ also reduces to the one for undeformed field theory which lives on the boundary of AdS black brane background.

So far we have discussed how the entanglement entropy behaves in different regimes of parameters in field theory. In the next section we will discuss the mutual information between two subsystems in a deformed field theory at non-zero temperature. 
	
\section{Holographic mutual information}\label{sec04}
As mentioned previously, entanglement entropy is a quantum measure used to study the correlations between two subsystems in a quantum state. However, for two disjoint intervals, A and B, where $A\cup B \subset \mathcal{C}$ and $\rho_{A\cup B}$ is not pure an important quantity to study total correlations (both classical and quantum correlations) between A and B is the mutual information (MI) \cite{r05}. It has been shown that mutual information follows an area law behaviour even at non-zero temperature \cite{r05}. In the holographic context it has been discussed that mutual information in field theories with holographic dual is monogamous which suggests that quantum entanglement dominates over classical correlations in such systems \cite{r08}. Mutual information is a linear combination of the entanglement entropy of the intervals and given by
\begin{equation}\label{DHMI}
  I(A:B)=S(A) +S(B) -S(A\cup B)~,
\end{equation}
where $S(A\cup B)$ is the entanglement entropy of the composite region $\rho_{A\cup B}$. This is a scheme independent quantity which is finite although the entanglement entropies have a divergent piece proportional to area. Also, sub-additivity guarantees that $I(A:B) \geq 0$ that is holographic mutual information is positive definite and equality is true when two disjoint regions are separable $\rho_{A\cup B}=\rho_{A}\otimes \rho_{B}$ \cite{Headrick:2010zt}.
\begin{figure}[hb]
  \centering
  \includegraphics[scale=0.7]{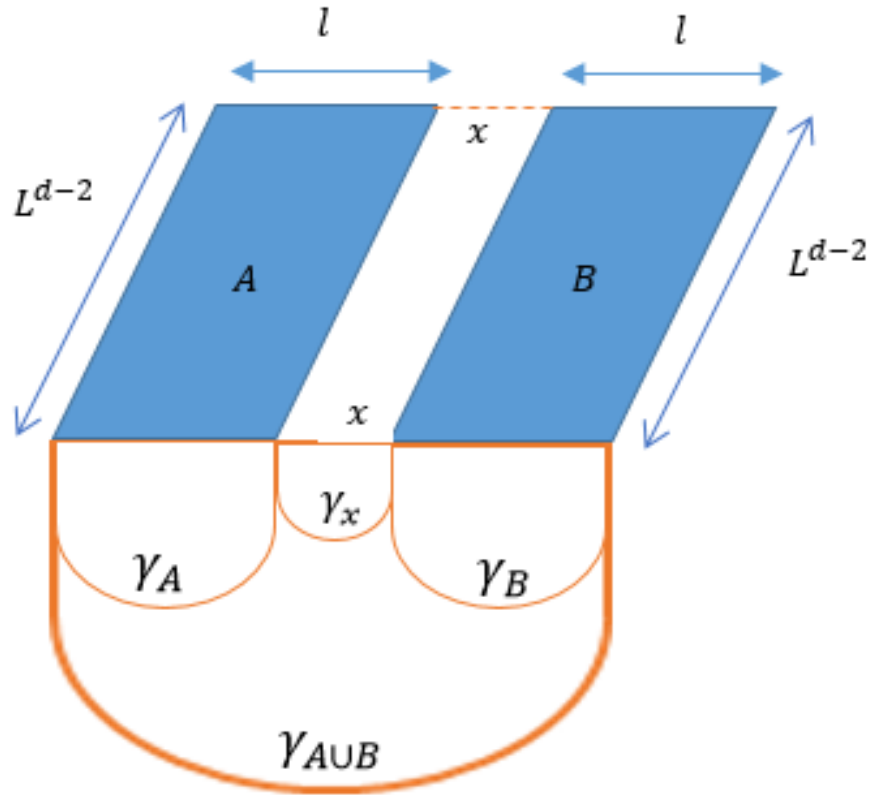}
  \caption{The schematic diagram of the two disjoint subsystems A and B, separated by a distance x with minimal area surfaces which is relevant for computing $S_{A\cup B}$ when the separation distance is small enough.}
  \label{fig:mutualinformation}
\end{figure}

To obtain mutual information in our set-up we consider two disjoint rectangular strips A and B, each of length $\ell$ in direction $x^1$ and L in other directions. They are separated with distance $x$ as in fig \ref{fig:mutualinformation}. In mutual information relation (\ref{DHMI}), holographic entanglement entropy for regions A and B are obtained by the usual Ryu-Takayanagi recipe. But for the composite region there are two options for minimal area surface in the bulk. If the separation distance is large  ($x/\ell\gg 1$) thus $S(A\cup B)=S(A)+S(B)=2 S(l)$ and $I(A:B) = 0$ which means the two subsystems are disentangled. For small separation distance ($x/\ell \ll 1$), we have $S(A\cup B)=S(2l+x)+ S(x)$ which produces a non-zero mutual information
\begin{equation}\label{HMI}
  I(A:B) = 2 S(l)- S(2l+x)- S(x)~.
\end{equation}
This means that one can obtain a critical separation distance as an upper bound for non-vanishing mutual information \cite{Headrick:2010zt}. Similar argument holds in the finite temperature case where the transition between zero and non-zero mutual information happens at a constant value of $Tx$ \cite{Molina-Vilaplana:2011ydi,r06}. This value depends on the dimension and temperature of the field theory. In our set-up in addition to temperature there is another scale in field theory which is the deformation parameter $\lambda$ of mass dimension $-d$. Working with $\tilde{\lambda}=\lambda^{1/d}$, another dimensionless combination appears in mutual information that is $\frac{x}{\tilde{\lambda}}$. This, in fact, will tell us how the dual gravity theory cut-off or deformation in field theory will affect the mutual information. As mentioned in the previous section, to be able to compare our results with the known results on entanglement entropy and mutual information we can work in the limit where $z_c$ or $\tilde{\lambda}$ is considered to be small. The detailed limits which enables us to use the analytical techniques introduced in \cite{Fischler:2012ca,r06} are discussed as follows.
		
In the last section we obtained the entanglement entropy for different limits of temperature and deformation parameter. To obtain the mutual information one needs to incorporate the distance between entangling regions, $x$, into these limits. As we mentioned earlier there already exists limits of $x$ where for $x \gg \ell$ and $x \gg 1/T$ the two subsystems disentangle and mutual information becomes zero \cite{r06}.  We assume such behaviour persists in our set-up and mutual information for $x/\ell \gg 1$ and $x T\gg 1$ vanishes.

Since we are interested in analytic results for mutual information we will work in the limits where we have already obtained the analytic entanglement entropy. These limits should be consistent with the limits on $x$ to get non-vanishing mutual information. Such consistency guides us to the following regimes for non-zero mutual information: small deformation regime where $\tilde{\lambda} \ll x \ll \ell$, large deformation regime as $x \ll \ell \ll \tilde{\lambda}$ and intermediate deformation regime where $x \ll \tilde{\lambda} \ll \ell $. Each of these contains different limits on the temperature regarding $\ell$ and $x$ which leads to five different results for mutual information as will be discussed in the following subsections. Just note that all these regimes have $\tilde{\lambda} T \ll 1$ in common. In gravity language it corresponds to $z_c/z_H \ll 1$ and comes from the bound that horizon puts on the cut-off. In fact specifying the limits helps us determine which result for entanglement entropy should be used for $S(x)$ and $S(\ell)$. Obviously $S(2 \ell+x)$ will follow the choice for $S(\ell)$.

\subsection{small deformation regime; \textbf{$\tilde{\lambda} \ll x \ll \ell$}}\label{subsmalldef}

In this regime where both $\tilde{\lambda}/x$ and $\tilde{\lambda}/\ell$ are smaller than one, we can use the results for entanglement in the small deformation regime for both $S(\ell)$ and $S(x)$ and consequently for $S(2\ell + x)$. In this regime we can incorporate two different temperature limits which can produce non-zero results for the mutual information, $T \ll 1/x , 1/\ell$ and $1/\ell \ll T \ll 1/x$. They correspond to small and intermediate regimes of temperature, respectively. We discuss each part individually.

\subsubsection{small deformation and small temperature regime; \textbf{$\tilde{\lambda} \ll x \ll \ell$} and \textbf{$T \ll 1/x , 1/\ell$}}

As it is obvious, for this regime we can use the result for entanglement entropy for low temperature and small deformation limit in subsection \ref{low small}. Therefore the relation (\ref{firstlimit}) can be used for $S(\ell)$, $S(x)$ and $S(2 \ell + x)$. We can expect this regime leads to the known result for mutual information in the $z_c \to 0$ or $\tilde{\lambda}\to 0$ limit as we work in small deformation regime. Using relation (\ref{firstlimit}) we get
\begin{equation}
\begin{aligned}\label{mutual1}
I(A:B) = \frac{2R^{d-1}}{4G_N^{(d+1)}}~\left(\frac{L}{z_c}\right)^{d-2}\Bigg\{&\left(a_1+ a_5 \left(\frac{z_c}{z_H}\right)^d+a_9 \left(\frac{z_c}{z_H}\right)^{2d}\right) \mathcal{F}_{d-2}
  + \left(a_2+a_6 \left(\frac{z_c}{z_H}\right)^{d}\right) \mathcal{F}_{2d-2}
  + a_3~ \mathcal{F}_{3d-2}\Bigg\}\\
  +\frac{2R^{d-1}}{4G_N^{(d+1)}}~\left(\frac{L}{z_H}\right)^{d-2}\Bigg\{&\left(a_4+a_8 \left(\frac{z_c}{z_H}\right)^d\right) \mathcal{G}_2
  + a_7 ~\mathcal{G}_{d+2}\Bigg\}~,
    \end{aligned}
    \end{equation}
where we have defined $\mathcal{F}_n \equiv z_c^n \left(\frac{2}{\ell^n}- \frac{1}{(2\ell+x)^n} - \frac{1}{x^n}\right)$ and $\mathcal{G}_n \equiv z_H^{-n} \left(2\ell^n - (2\ell+x)^n - x^n\right)$. We just remind the reader that the $a_i$ coefficients are just numbers depending on the dimension of field theory $d$ defined in relations (\ref{as}). The sign of these coefficients does not depend on dimension.

Let's check whether this result meets the conditions that mutual information in field theories with holographic dual should satisfy. As one can see there is no term in the mutual information with $z_c$ in the denominator and therefore setting $z_c$ to zero we do not get any divergences as one expects. This is due to the fact that we are working in the small deformation regime and we are allowed to set $z_c$ to zero in this result. We can also observe that the mutual information remains positive as long as we have $x$ smaller than a critical value which depends on temperature and deformation parameter. Therefore the corrections due to both temperature and deformation respect the positive-definiteness of mutual information. Note that increasing $x$ leads to decreasing the mutual information until it becomes zero and the two subsystems disentangle. The value of $x$ where mutual information becomes zero is the critical value where the phase transition between zero and non-zero correlation happens \cite{Headrick:2010zt,r06}. This critical value depends on $\lambda$ besides depending on $\ell$ and $T$. Our result also shows that although the corrections due to temperature get modified by corrections due to $z_c$ but the mutual information decreases by corrections due to both temperature and deformation parameter. In fact the sign of the sum of all the corrections is negative and reduces the mutual information. Therefore we have shown that corrections due to deformation parameter or cut-off combined with corrections due to temperature decrease the correlation between two subsystems in deformed field theory.

Since we are working in the small deformation regime we can check whether our result leads to the known results for the mutual information in an undeformed field theory. We set the limit $z_c \to 0$ and this gives us the mutual information corrected only by the finite temperature. In this limit (\ref{mutual1}) reduces to
 \begin{equation}
 \begin{aligned}\label{mutual1T}
 I(A:B)|_{z_c\to 0} =&\frac{2R^{d-1}}{4G_N^{(d+1)}}~L^{d-2}\Bigg\{a_1 \left(\frac{2}{\ell^{d-2}}- \frac{1}{(2\ell+x)^{d-2}} - \frac{1}{x^{d-2}}\right)\\
 +&\frac{2R^{d-1}}{4G_N^{(d+1)}}~L^{d-2}\bigg\{-2 \frac{a_4}{z_H^d} \left(\ell+x\right)^2
  +\frac{a_7}{z_H^{2d}} \left(2\ell^{d+2} - (2\ell+x)^{d+2} - x^{d+2}\right)\bigg\}~.
    \end{aligned}
    \end{equation}
The first line in this result is exactly the mutual information between two subsystems in the field theory dual to $AdS_{d+1}$ \cite{r06}
 \begin{equation}
 I(A:B)|_{T\to 0, z_c\to 0} =  \frac{2R^{d-1}}{4G_N^{(d+1)}}~L^{d-2} a_1 \left(\frac{2}{\ell^{d-2}}- \frac{1}{(2\ell+x)^{d-2}} - \frac{1}{x^{d-2}}\right)~.
 \end{equation}
The result we have here is up to $2 d$ order in temperature. Comparing our result with what is known in the literature which includes  only terms up to order $d$ in temperature \cite{r06} we see that they exactly match. The $T^d$ order term is negative while the other correction, term of order $T^{2d}$ is positive. Since this result is a perturbative calculation higher order terms have smaller effect and therefore the combined result is always negative. Therefore the corrections due to temperature alone decrease the mutual information in the field theory as mentioned in \cite{r06}. We see here that even in the presence of higher order corrections this result remains valid. In fact comparing the leading and subleading correction terms leads to an upper value for $\ell/z_H$ or $\ell T$ where the perturbation breaks down and the result here is no more reliable.

Now we can check what our result for mutual information tells us about the correlation between two subsystems when it is only corrected by deformation parameter. So we can set $T\to 0$ or $z_H \to \infty$ in (\ref{mutual1}) while we keep $z_c$ a finite value. In this limit all the terms proportional to $\mathcal{G}_n$ and $z_c/z_H$ in (\ref{mutual1})  vanish and we are left with
\begin{equation}
\begin{aligned}\label{mutual1zc}
 I(A:B)|_{T\to 0} &= I(A:B)|_{T\to 0, z_c\to 0}\\
  +&\frac{2R^{d-1}}{4G_N^{(d+1)}}~L^{d-2}\Bigg\{ a_2 z_c^d \left(\frac{2}{\ell^{2 d-2}}- \frac{1}{(2\ell+x)^{2 d-2}} - \frac{1}{x^{2 d-2}}\right)+ a_3~z_c^{2d} \left(\frac{2}{\ell^{3d-2}}- \frac{1}{(2\ell+x)^{3d-2}} - \frac{1}{x^{3d-2}}\right)\Bigg\}~,
 \end{aligned}
 \end{equation}
 which, up to order $z_c^d$ matches the result in \cite{Khoeini-Moghaddam:2020ymm} for the geometry $AdS_{d+1}$ with cut-off. Note that similar to temperature the correction terms in the above relation are all proportional to the area of the subsystems $L^{d-2}$. Therefore we can conclude that, in the small deformation regime, the corrections in field theory due to the deformation parameter $\tilde{\lambda}$ are proportional to the area of the entangling region. Another interesting observation is that, similar to temperature, the first correction term in the above result which is of order $z_c^d$ is negative while the other term which is of order $z_c^{2d}$ is positive but almost zero. The combined result of the corrections in $z_c$ or deformation parameter is negative. Therefore we can see that in zero temperature limit the corrections due to the deformation reduces the mutual information in deformed field theory. It's interesting to conclude that the corrections due to the deformation parameter in field theory or cut-off in gravity behaves similarly to the corrections due to temperature. This confirms our conclusion made previously in the entanglement entropy.

 Another interesting observation is that, in the small deformation regime, even though there are terms proportional to $z_c$, if we set the distance between the subregions $x$ to zero we get divergent mutual information. This means that in the small deformation regime the presence of $z_c$ can not diminish the divergence in mutual information due to $x\to 0$. We will see later on that in the large deformation regime this divergence does not appear.

\begin{figure}[hp]
  \includegraphics[scale=0.64]{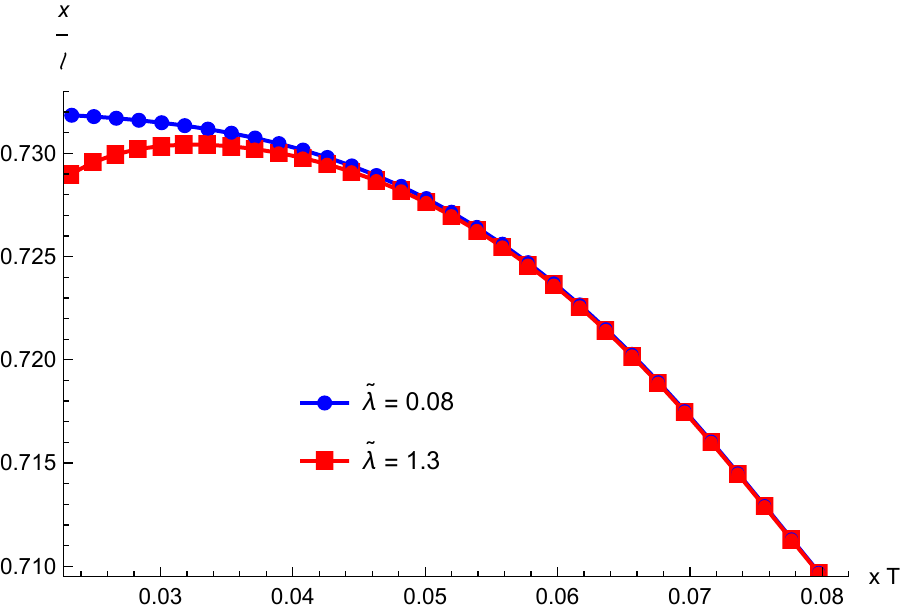}
 \includegraphics[scale=0.64]{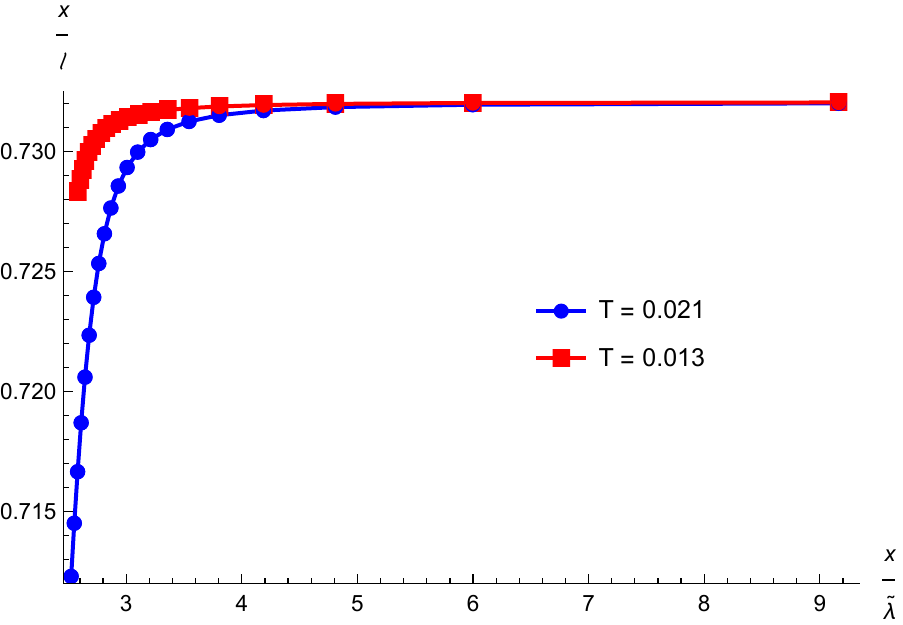}
  \includegraphics[scale=0.64]{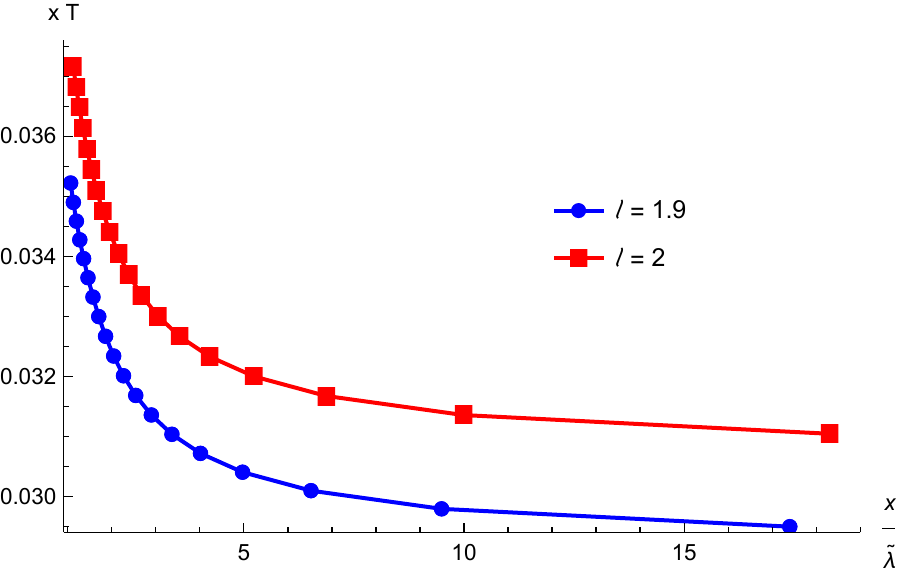}
  \caption{\textbf{Left:} $x/\ell$ is plotted with respect to $xT$ for two different values of deformation parameter $\tilde{\lambda}$. The range of parameters in this limit are chosen as $0.03 < \ell T <0.11$, $\ell/\tilde{\lambda} > 2.35$ and $\tilde{\lambda} T < 0.016$. \textbf{Middle:} $x/\ell$ is plotted with respect to $x/\tilde{\lambda}$ for two different values of temperature $T$. The range of parameters in this limit are chosen as $0.01 < \ell T <0.1$, $\ell/\tilde{\lambda} > 3.5$ and $\tilde{\lambda} T < 0.03$. \textbf{Right:}  $x T$ is plotted with respect to $x/\tilde{\lambda}$ for two different values of entangling length $\ell$. The range of parameters in this limit are chosen as $0.04 < \ell T <0.52$, $\ell/\tilde{\lambda} > 1.5$ and $\tilde{\lambda} T < 0.04$. Note that we have set $R=G_N=1$. }
\label{fig:ma}
\end{figure}

To better see how the phase transition between zero and non-zero mutual information is affected by the presence of temperature and deformation in field theory we have plotted the phase transition diagrams both with respect to temperature and deformation parameter in figure \ref{fig:ma}.
 In this figure we have plotted $x/\ell$ with respect to $xT$ (left), $x/\tilde{\lambda}$ (middle) and $xT$ with respect to $x/\tilde{\lambda}$ (right). The range of parameters under each graph will produce non-zero mutual information. Each plot shows that the phase transition between zero and non-zero mutual information depends on the field theory parameters $T,~\lambda,~\ell$ and the dimension of the field theory $d$. The first plot, figure \ref{fig:ma}-left, tells us that there exists an upper limit on the value of both $x/\ell$ and $xT$ ratios where the two subsystems decouple, as we expect. We can also see that as the temperature increases the phase transition happens at smaller values of the ratio $x/\ell$. Note that moving to the right on the horizontal axis corresponds to increasing the temperature. This means that for states at higher temperature the two subsystems decouple at smaller values of separation distance $x$. It is consistent with the result discussed previously that correction terms due to temperature decreases the mutual information or correlation between two subsystems. This figure also tells us that the deformation parameter can help the temperature and reduce the value of $x$ at which the decoupling of the two subsystems happens even more. For the states at smaller temperature the effect of the deformation parameter is stronger than higher temperature states. For two subsystems with the same size and at the same temperature but different values of deformation parameter, the mutual information becomes zero at smaller values of $x/\ell$ if the deformation parameter is larger. Therefore this plot suggests that, in the regime of parameters we work in this section, the effect of temperature will be more dominant than deformation parameter on disentangling the two subsystems as the temperature is raised.

The middle plot in figure \ref{fig:ma} also shows an upper limit on $x/\ell$ but, in contrast to $xT$, it gives a lower bound on $x/\tilde{\lambda}$ as the range of parameters in this section suggests. In this figure moving to the right on the horizontal axis corresponds to decreasing the deformation parameter $\tilde{\lambda}$. The figure tells us that as the deformation parameter increases the phase transition happens at smaller values of $x/\ell$. This is similar to the dependence of the decoupling of the two subsystems on the temperature discussed in the right plot. This means that the two subsystems can disentangle in a deformed theory at smaller separation distances compared to undeformed theory. It also suggests that at smaller values of deformation parameter the decoupling of the subsystems for states at different temperatures depends dominantly on $\ell$ not temperature. But as the deformation parameter is raised the effect of temperature becomes more recognizable. Therefore for subsystems with the same size and the same deformation parameter the subsystems disentangle at shorter separation distances for states at higher temperature.

The last plot in figure \ref{fig:ma} (the one on the right) gives us an upper bound on the ratio $\tilde{\lambda} T$. It means that to have non-zero mutual information or correlation between two subsystems, the deformation parameter should be smaller than some numerical value over temperature. In the dual gravity this confirms that the cut-off radius $z_c$ should be always smaller than the horizon radius $z_H$ times a numerical value. In fact it tells us that $z_H$ can not get arbitrarily close to the cut-off as the range of parameters we work in is not valid anymore. One should note that this numerical value strongly depends on the entangling length $\ell$. It also suggests that for larger subsystems (larger $\ell$) the disentangling happens at larger values of the combination $\tilde{\lambda} T$ or in gravity dual larger values of $z_c/z_H$. This is consistent with what we expect since larger $\ell$ in field theory means surfaces deeper into the bulk. Therefore the two subsystems disentangle more easily if $z_c$ approaches $z_H$.

\subsubsection{small deformation and intermediate temperature regime; \textbf{$\tilde{\lambda} \ll x \ll \ell$} and \textbf{$1/\ell \ll T \ll 1/x$}}

The difference between this section and the previous one is in the fact that we are working in the high temperature and small deformation regime for the entangling length $\ell$ but low temperature and small deformation regime for the distance $x$. Therefore we have to use the entanglement entropy obtained in subsection \ref{high small} relation (\ref{thirdlimit}) for $S(\ell)$ and $S(2 \ell +x)$ and the relation (\ref{firstlimit}) for $S(x)$. The result for mutual information is
%
%
\begin{equation}
\begin{aligned}
I(A:B) =\frac{2R^{d-1}}{4G_N^{(d+1)}}\left(\frac{L}{z_H}\right)^{d-2}&\Biggl{\{}\hat{\mathcal{S}}-\sqrt{\frac{d-1}{2d}}\left(2 \epsilon_\ell - \epsilon_{2\ell+x}\right)+\frac{1}{2d}\left(\frac{z_c}{z_H}\right)^d+\frac{1}{8d}\left(\frac{z_c}{z_H}\right)^{2d}\\
&-\frac{x}{2z_H}-\left(a_4+a_8 \left(\frac{z_c}{z_H}\right)^d\right) \left(\frac{x}{z_H}\right)^2- a_7 \left(\frac{x}{z_H}\right)^{d+2} \Biggr{\}}\\
-\frac{2R^{d-1}}{4G_N^{(d+1)}}\left(\frac{L}{x}\right)^{d-2}&\Biggl{\{}\left(a_1+ a_5 \left(\frac{z_c}{z_H}\right)^d+a_9 \left(\frac{z_c}{z_H}\right)^{2d}\right)+\left(a_2+a_6 \left(\frac{z_c}{z_H}\right)^{d}\right) \left(\frac{z_c}{x}\right)^d
+a_3 \left(\frac{z_c}{x}\right)^{2d} \Biggr{\}}~,
\end{aligned}
\end{equation}
where $\epsilon_\ell$ and $\epsilon_{2\ell+x}$ means the quantity $\epsilon$ in (\ref{epsilon}) evaluated at $\ell$ and $2\ell+x$,  respectively. We again observe that all the corrections to mutual information due to the deformation in field theory follow the area law behaviour. This result also shows that the corrections due to deformation or cut-off in the mutual information appear in two ways: the terms that are proportional to the ratio $(z_c/z_H)^{nd}$ and only depend on dimension versus the terms that depend on the separation distance between the subsystems. The first set of terms are constant for a deformed theory at a specific temperature but, due to $\hat{\cal{S}}$, give negative contribution to the mutual information. These terms depend on the parameters of the theory and not on the scale of the non-local operator we calculate. These terms are present in the mutual information in this regime of parameters no matter what the separation between subsystems, $x$, is.  The terms involving $x$ are the ones that produce positive contribution to the mutual information and lead to non-zero correlation between subsystems. Note that like the previous limit if we send the separation distance to zero the mutual information diverges. The small deformation limit can not regulate this divergence. 

Since we are working in the large temperature limit for $T\ell$ the relevant check here is to send $z_c$ to zero and see whether we get the results already obtained in the literature \cite{r06}. In this limit we get
\begin{equation}
I(A:B)|_{z_c \to 0} =\frac{2R^{d-1}}{4G_N^{(d+1)}}\left(\frac{L}{z_H}\right)^{d-2}\biggl{\{}\hat{\mathcal{S}}-a_1 \left(\frac{z_H}{x}\right)^{d-2}-\frac{x}{2z_H} - a_4 \left(\frac{x}{z_H}\right)^2
  - a_7 \left(\frac{x}{z_H}\right)^{d+2} \biggr{\}}~,
\end{equation}
where we have assumed the $\epsilon$ terms go to zero in the large $T\ell$ limit. This result exactly matches the result for mutual information in the field theory dual to $AdS_{d+1}$ black brane background. 

As a concluding remark for the regime of small deformation we observe that the result of these two subsections have individual correction terms in temperature and in deformation parameter in addition to terms proportional to $(\tilde{\lambda} T)^d$. As we will see such behaviour does not necessarily persist in the other limits.  

\subsection{large deformation regime \textbf{$x \ll \ell \ll \tilde{\lambda}$}}\label{sublargedef}

Interestingly in this regime, the consistency of the limits forces us to have only low temperature regime for both $\ell$ and $x$. Therefore we have to work in the limit where $T \ll 1/x , 1/\ell$. As a result we have to use the relation for the entanglement entropy obtained in section \ref{low large} equation (\ref{secondlimit}) for $S(\ell)$, $S(x)$ and consequently $S(2 \ell+x)$.
The mutual information in this limit becomes
\begin{equation}
\begin{aligned}\label{HMI2}
  I(A:B)=&\frac{R^{d-1}}{4G_N^{(d+1)}}\left(\frac{L}{z_c}\right)^{d-2}\Biggl{\{}  -\frac{2x}{z_{c}}-\frac{(d-1)^2}{24}\left(1-\left(\frac{z_c}{z_H}\right)^d\right)~\tilde{\mathcal{G}}_{3}\\
  +&\frac{(d-1)^3}{1920}(d+7)
\left(1+ 2 \frac{d-7}{d+7}\left(\frac{z_c}{z_H}\right)^d-\frac{3d-7}{d+7}\left(\frac{z_c}{z_H}\right)^{2d}\right)~\tilde{\mathcal{G}}_5 \Biggr{\}}~,
\end{aligned}\end{equation}
where $\tilde{\mathcal{G}}_n \equiv z_c^{-n} \left(2\ell^n - (2\ell+x)^n - x^n\right)$. This function is similar to $\mathcal{G}_n$ introduced in the previous section except for $z_H$ which is replaced by $z_c$. This result shows that, in contrast to the zero cut-off and small deformation cases, in the limit of zero separation distance ($x \rightarrow 0$) mutual information remains finite and does not diverge although we are working in the small temperature limit. The mutual information is proportional to the area of the subsystem and even cut-off presence does not change it.  Also we can observe that for a fixed cut-off when the separation distance increases the mutual information decreases as one expects. 

Another point in the above result is that the dependence on temperature in mutual information appears only as $nd$ orders of the dimensionless parameter $z_c/z_H$ or $\tilde{\lambda}T$ while the cut-off dependence is different. It seems that in the large deformation regime in the deformed field theory the temperature does not affect the mutual information independently and its effect is always through the product $\tilde{\lambda}T$. Also note that in the above result the dependence on subsystem scale or entangling length $\ell$ is through terms that are universal and the orders of $\ell$ do not depend on the dimension of the theory. 
\begin{figure}[ht]
  \includegraphics[scale=0.64]{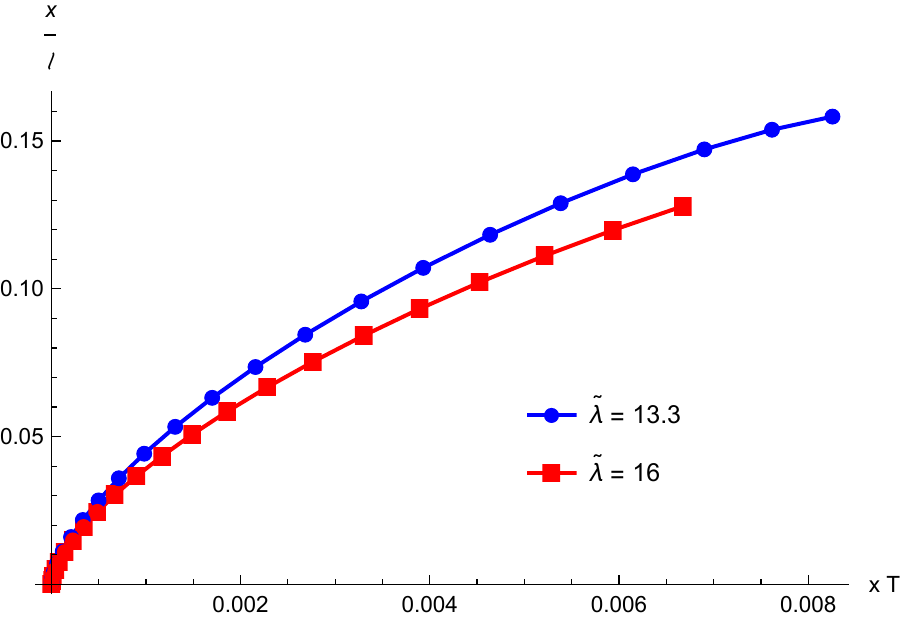}
 \includegraphics[scale=0.64]{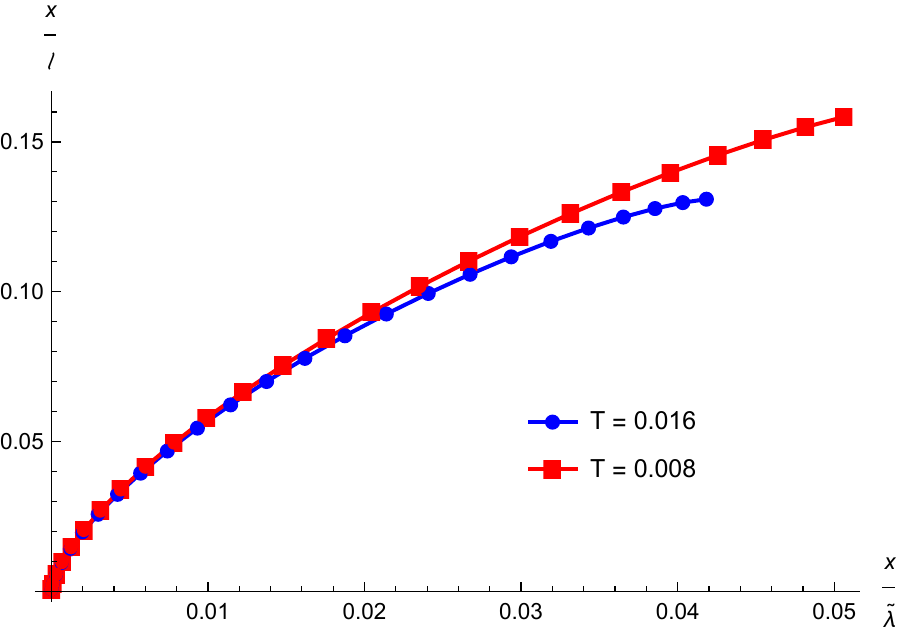}
  \includegraphics[scale=0.64]{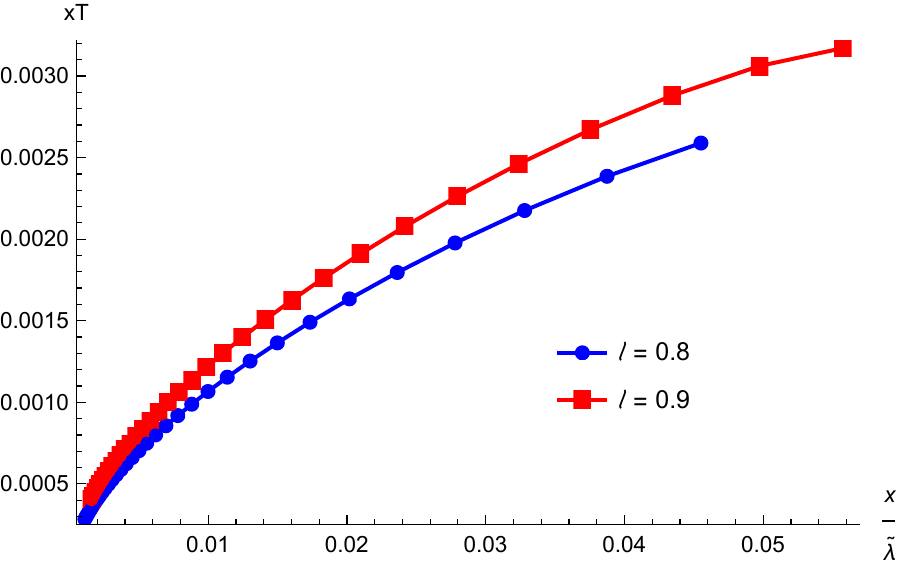}
  \caption{\textbf{Left:} $x/\ell$ is plotted with respect to $xT$ for two different values of deformation parameter $\tilde{\lambda}$. The range of parameters in this limit are chosen as $\ell T < 0.052$, $\ell/\tilde{\lambda} < 0.33$ and $\tilde{\lambda} T < 0.2$. \textbf{Middle:} $x/\ell$ is plotted with respect to $x/\tilde{\lambda}$ for two different values of temperature $T$. The range of parameters in this limit are chosen as $0.002 < \ell T <0.08$, $\ell/\tilde{\lambda} < 0.32$ and $\tilde{\lambda} T < 0.25$. \textbf{Right:}  $x T$ is plotted with respect to $x/\tilde{\lambda}$ for two different values of entangling length $\ell$. The range of parameters in this limit are chosen as $0.017 < \ell T <0.025$, $\ell/\tilde{\lambda} < 0.34$ and $\tilde{\lambda} T < 0.25$. Note that we have set $R=G_N=1$. }
\label{fig:mb}
\end{figure}

The phase transition diagrams for the mutual information in this limit have been plotted in figure \ref{fig:mb}. The first one on the left is $x/\ell$ plotted with respect to $xT$. Such behaviour is similar to the graph in the middle where $x/\ell$ is plotted with respect to $x/\tilde{\lambda}$. This similarity comes from the discussion we had about temperature dependence in the last paragraph. If we plot these diagrams for fixed $\ell$ we observe that $x/\ell$ decreases by both increasing the temperature and deformation parameter. Therefore raising the temperature or deformation parameter in field theory causes the phase transition to happen at smaller values of separation distance. This means that at higher values of temperature and deformation parameter the correlation between two subsystems decreases. 

Figure \ref{fig:mb}-left shows that for small values of $T$ and $\ell$ the ratio $x/\ell$ is not sensitive to the deformation parameter. But as temperature is raised the larger deformation parameter results in less correlated subsystems and mutual information becomes zero for smaller values of $x/\ell$. Note that in this plot moving to the right on the horizontal axis corresponds to raising the temperature. The same thing happens  in figure \ref{fig:mb}-middle replacing $T$ with deformation parameter $\tilde{\lambda}$. For small values of $\tilde{\lambda}$ and $\ell$ the ratio $x/\ell$ is not affected by $T$. But for larger values of deformation parameter the mutual information differs between different $T$s and mutual information between to subsystems becomes zero at smaller separation distances for larger values of $T$. Also in this plot moving to the right means increasing the deformation parameter. 

In the last plot in figure \ref{fig:mb}, moving towards right direction is decrease in $T$ and $\tilde{\lambda}$. Therefore for smaller values of temperature and deformation parameter in field theory the scale of the subsystem in correlation between the two becomes more effective. 

Since we are working in the small temperature regime we can send $T\to 0$ or $z_H\to \infty$ and we get
\begin{equation}
  I(A:B)|_{T \to 0}=\frac{R^{d-1}}{4G_N^{(d+1)}}\left(\frac{L}{z_c}\right)^{d-2}\biggl{\{}  -\frac{2x}{z_{c}}-\frac{(d-1)^2}{24}~\tilde{\mathcal{G}}_{3}
  +\frac{(d-1)^3}{1920}(d+7)~\tilde{\mathcal{G}}_5 \biggr{\}}~.
  \end{equation}
We can also solve the equation for mutual information (\ref{HMI2}) for  $x$ perturbatively. This is where the mutual information becomes zero and in other words the phase transition between zero and non-zero mutual information happens. We obtain
\begin{equation}
\frac{x}{z_c} = \frac{(d-1)^2}{8} \left(1-\left(\frac{z_c}{z_H}\right)^d\right) \left(\frac{\ell}{z_c}\right)^3-\frac{(d-1)^3}{128} \left(1-\left(\frac{z_c}{z_H}\right)^d\right) \left(11-3d+\left(7d-11\right)\left(\frac{z_c}{z_H}\right)^d\right)\left(\frac{\ell}{z_c}\right)^5 + ... ~,
\end{equation}
where the dots are terms of order $(\ell/z_c)^7$ with terms up to order $(z_c/z_H)^{3 d}$. It is interesting to see that the ratio $x/z_c$ similar to $x/\ell$ can define a value for phase transition which depends on temperature $T$, $\ell$ and deformation parameter $\tilde{\lambda}$. In th next section we consider intermediate deformation regime and discuss how the mutual information differs from the other limits.

\subsection{intermediate deformation regime \textbf{$x \ll \tilde{\lambda} \ll \ell$}}\label{subinterdef}

Similar to the first regime discussed in the mutual information section, this regime accepts two different limits for temperature, small temperature regime where ($T \ll 1/x , 1/\ell$) and intermediate temperature one where ($1/\ell \ll T \ll 1/x$). We discuss these limits separately as follows.

\subsubsection{intermediate deformation and small temperature regime; \textbf{$x \ll \tilde{\lambda} \ll \ell$} and \textbf{$T \ll 1/x , 1/\ell$}}
\label{subC1}

In this regime parameters $\ell$ and $x$ are treated differently with respect to deformation parameter. Therefore we have to use low temperature and small deformation result for entanglement entropy, relation (\ref{firstlimit}) for $S(\ell)$ and $S(2\ell +x)$. For $S(x)$ we need to use low temperature and large deformation limit for entanglement entropy, relation (\ref{secondlimit}).

For this limit replacing equations \eqref{firstlimit} and \eqref{secondlimit} into \eqref{HMI} we are left with the following result for the holographic mutual information
\begin{equation}\label{MI1}
  \begin{aligned}
 I(A:B) =\frac{2R^{d-1}}{4G_N^{(d+1)}}\left(\frac{L}{z_c}\right)^{d-2}&\Biggl{\{}\frac{1}{d-2}-\frac{1}{4}\left(\frac{z_c}{z_H}\right)^{d}-\frac{3}{8(d+2)} \left(\frac{z_c}{z_H}\right)^{2d}-\left(\frac{x}{2z_{c}}\right)+\frac{(d-1)^2}{6}\left(1-\left(\frac{z_c}{z_H}\right)^{d}\right)\left(\frac{x}{2z_{c}}\right)^{3}\\
 &-\frac{(d-1)^3}{120}(d+7)
\left(1+ 2 \frac{d-7}{d+7}\left(\frac{z_c}{z_H}\right)^d-\frac{3d-7}{d+7}\left(\frac{z_c}{z_H}\right)^{2d}\right)~\left(\frac{x}{2 z_c}\right)^5\\
  &+\left(a_1 + a_5 \left(\frac{z_c}{z_H}\right)^d + a_9 \left(\frac{z_c}{z_H}\right)^{2d}\right)\hat{\mathcal{F}}_{d-2}
  +\left(a_2 + a_6 \left(\frac{z_c}{z_H}\right)^d\right)\hat{\mathcal{F}}_{2d-2}+ a_3~ \hat{\mathcal{F}}_{3d-2}\Biggr{\}}\\
 +\frac{2R^{d-1}}{4G_N^{(d+1)}}\left(\frac{L}{z_H}\right)^{d-2}&\Biggl{\{}\left(a_4 + a_8 \left(\frac{z_c}{z_H}\right)^d\right) \hat{\mathcal{G}}_{2} + a_7~ \hat{\mathcal{G}}_{d+2}\Biggr{\}}~,
\end{aligned}
\end{equation}
where $\hat{\mathcal{F}}_{n}= z_c^n \left(\frac{2}{\ell^n}- \frac{1}{(2\ell+x)^n}\right)$ and $\hat{\mathcal{G}}_{n}=z_H^{-n} \left(2l^n - (2l+x)^n\right)$ are in fact $\mathcal{F}_n$ and $\mathcal{G}_n$ without the $x^{-n}$ and $x^n$ terms, respectively. Similar to small deformation regime we have corrections to mutual information coming from temperature and deformation parameter individually. Both temperature and deformation correction terms decrease the mutual information. But similar to large deformation regime the mutual information does not diverge if the separation distance between two subsystems goes to zero. 

Since we are working in low temperature limit one can send $T\to 0$ and see the specific effect of deformation parameter at some finite cut-off. Doing so we get the following result for mutual information
\begin{equation}\label{MI1}
  \begin{aligned}
 I(A:B)|_{T \to 0} =\frac{2R^{d-1}}{4G_N^{(d+1)}}\left(\frac{L}{z_c}\right)^{d-2}&\Biggl{\{}\frac{1}{d-2}-\left(\frac{x}{2z_{c}}\right)+\frac{(d-1)^2}{6}\left(\frac{x}{2z_{c}}\right)^{3}
 -\frac{(d-1)^3}{120}(d+7)\left(\frac{x}{2 z_c}\right)^5\\
  &+a_1~\hat{\mathcal{F}}_{d-2}+a_2~\hat{\mathcal{F}}_{2d-2}+ a_3~ \hat{\mathcal{F}}_{3d-2}\Biggr{\}}~.
\end{aligned}
\end{equation}
Interestingly there are two expansion parameters in this result; one is $x/z_c$ correction terms on the first line and the other one is $x/\ell$ correction terms coming from terms on the second line. Therefore it can be clearly seen that the critical separation for transition between non-zero and zero mutual information depends not only on $\ell$ but also on $z_c$ or $\tilde{\lambda}$. This result again confirms that increasing deformation parameter reduces the mutual information. So the disentanglement between the two subsystems happens at smaller separation distance compared to undeformed theory. Increasing $z_c$ or deformation parameter, one should not forget that there exists an upper limit $z_c\ll \ell$ on $z_c$ to obtain a reliable result.

\subsubsection{intermediate deformation and intermediate temperature regime; \textbf{$x \ll \tilde{\lambda} \ll \ell$} and \textbf{$1/\ell \ll T \ll 1/x$}}

This regime imposes opposite limits on the entangling length $\ell$ and separation distance $x$. Therefore we use the result for entanglement entropy in low temperature and large deformation limit, relation (\ref{secondlimit}), for $S(x)$ and high temperature, small deformation limit entanglement entropy, relation (\ref{thirdlimit}) for $S(\ell)$ and $S(2\ell +x)$.

Substituting equations \eqref{thirdlimit} and \eqref{secondlimit} into the equation \eqref{HMI} gives the following relation for the mutual information in this limit
\begin{equation}
\begin{aligned}
  I(A:B)=\frac{2R^{d-1}}{4G_N^{(d+1)}}\left(\frac{L}{z_c}\right)^{d-2}&\Biggl{\{} \frac{1}{d-2}-\left(1+\left(\frac{z_c}{z_H}\right)^{d-1}\right)\left(\frac{x}{2z_c}\right)
  +\frac{(d-1)^2}{6}\left(1-\left(\frac{z_c}{z_H}\right)^d\right)\left(\frac{x}{2z_c}\right)^3 \\
  &-\frac{(d-1)^3}{120}(d+7)
\left(1+ 2 \frac{d-7}{d+7}\left(\frac{z_c}{z_H}\right)^d-\frac{3d-7}{d+7}\left(\frac{z_c}{z_H}\right)^{2d}\right)~\left(\frac{x}{2 z_c}\right)^5\\
  &-\frac{1}{4}\left(\frac{z_c}{z_H}\right)^d-\frac{3}{8(d+2)}\left(\frac{z_c}{z_H}\right)^{2d}
  \Biggr{\}}\\ 
  +\frac{2R^{d-1}}{4G_N^{(d+1)}}\left(\frac{L}{z_H}\right)^{d-2}&\Biggl{\{}\left(\hat{\mathcal{S}}-\sqrt{\frac{d-1}{2d}}\left(2\epsilon_{\ell} - \epsilon_{2\ell+x} \right)\right)+\frac{1}{2d}\left(\frac{z_c}{z_H}\right)^{d}+\frac{1}{8d}\left(\frac{z_c}{z_H}\right)^{2d}\Biggr{\}}~.
  \end{aligned}
\end{equation}
This result also shows that in the intermediate deformation regime mutual information does not diverge if we set the separation distance to zero. In this result, like the previous one in subsection \ref{subC1}, we have individual expansions in temperature and in deformation parameter. It seems that only in large deformation regime there is no expansion in temperature independent of deformation parameter. But in contrast to the previous result, in this limit the expansion parameter $x/\ell$ only appears through $\epsilon_{2\ell+x}$. If we consider $x \ll \ell$ we can ignore the effect of it and therefore the phase transition in the mutual information is defined by $x/z_c$.

	\section{Summary and concluding remarks}\label{summary}
	
In this work we have discussed different limiting behaviour of both entanglement entropy and mutual information in deformed field theory at finite temperature. We have showed that deformation parameter and temperature have similar effect on holographic entanglement entropy. At low temperature or small deformation limit the entanglement entropy is proportional to the area of the subsystem. It means that in this limit the degrees of freedom near the boundary of the subsystem are involved in the correlation between two subsystems. But at large deformation parameter or high temperature limit it behaves like thermal entropy and is proportional to the volume of the subsystem. Therefore it seems that increasing the temperature or deformation parameter causes other degrees of freedom to be included in the correlation between two subsystems. The deformation in field theory by generalized $T\bar{T}$ operator translates into introducing non-zero cut-off in the bulk. So high temperature or large deformation means probing deeper into the bulk and this seems to cause the thermal fluctuations to have the dominant effect. 

There are three dimensionless expansion parameters in the analytic result for the entanglement entropy, $\tilde{\lambda}/\ell$, $T \ell$ and $\tilde{\lambda} T$, where $T$, $\tilde{\lambda}$ and $\ell$ are temperature, deformation parameter which is proportional to cut-off and subsystem scale in field theory, respectively. The first two expansion parameters are due to the non-local operator with scale $\ell$ that we are studying in field theory and the last one is composed of only field theory scales and knows nothing about the non-local operator. The last expansion parameter consistently emerges from the calculations. An interesting observation in our results is that, in the regime of small temperature and deformation, all three dimensionless expansion parameters show up in the final result. But for large deformation or large temperature the expansion parameters reduce to two. In fact the effect of small parameter appears only through $\tilde{\lambda} T$ and it has no independent effect on the entanglement entropy. It seems that the non-local operator does only see the dominant effect due to larger parameter. 

We have also discussed the holographic mutual information in various regimes of parameters. One of the common features between all of these limits is that the mutual information follows the area law behaviour. The other one is that temperature and deformation correction terms reduce the mutual information and therefore the quantum correlation between the two subsystems. This is consistent with the result from entanglement entropy that temperature and deformation correction terms add negative contribution to the entanglement entropy. Therefore increasing the temperature and deformation parameter causes the mutual information to become zero at smaller separation distances. We have shown this result in various phase diagrams we have plotted for the mutual information in deformed field theory at finite temperature. 

One of the other results of our paper is that in the large or intermediate deformation regimes (defined in subsections \ref{subinterdef} and \ref{sublargedef}) the mutual information does not diverge if the separation distance between the two subsystems goes to zero. This shows that  deformation can correct the mutual information in a way that correlation between the two subsystems do not become infinite if they merge. This is not true in the small deformation regime, subsection \ref{subsmalldef}. Therefore how large the deformation is compared to the other scales in the theory is important in this discussion. Interestingly we observe that high temperature regime does not produce such result and only deformation parameter leads to it. 

Following this work it is interesting to see what the effect of chemical potential might be on entanglement entropy in a deformed theory. The deformation operator in field theory gets modified due to the conserved current corresponding to the chemical potential. It can be discussed whether chemical potential has similar effect on entanglement entropy to temperature and deformation parameter. Another question that the authors are currently involved with is whether the correlation function in holographic deformed field theory depends on deformation parameter and temperature the same way we observe in holographic calculations. Whether we can see the emergence of the dimensionless parameter $\tilde{\lambda}T$ there too. 

		\section*{Acknowledgments}
	H. E. would like to thank M. Ali-Akbari for fruitful discussions. This work is based upon research funded by Iran National Science Foundation (INSF) under Project No. 99024816.

	\appendix
	\section{Used Mathematical Relations}\label{app1}
	In this appendix we present some useful relations which has been used throughout the paper.
	\begin{enumerate}
		\item {\textbf{Newton's binomial and trinomial expansion}}\;\; \\\\
		Newton's generalized binomial expansion when $\lvert y\rvert<\lvert x\rvert$  is given by
		\begin{equation}
			\begin{aligned}
				\left(x+y\right)^r&=\sum_{k=0}^{\infty} \binom{r}{k}\,x^{r-k}\,y^k\,,\\
				\left(x+y\right)^{-r}&=\sum_{k=0}^{\infty}(-1)^k \binom{r+k-1}{k}\,x^{-r-k}\,y^k\,.
			\end{aligned}
		\end{equation}
		Similarly the generalized trinomial expansion for $\lvert y+z\rvert<\lvert x\rvert$ is given by
		\begin{equation}
			\begin{aligned}
				\left(x+y+z\right)^r&=\sum_{k=0}^{\infty}\sum_{j=0}^{k} \binom{r}{k}\,\binom{k}{j}\,x^{r-k}\,y^{k-j}\,z^j\,,\\
				\left(x+y+z\right)^{-r}&=\sum_{k=0}^{\infty}\sum_{j=0}^{k}(-1)^k \binom{r+k-1}{k}\,\binom{k}{j}\,x^{-r-k}\,y^{k-j}\,z^j\,,
			\end{aligned}
		\end{equation}
		where $x,\,y,\,r\in \mathbb{R}\,$ and $r>0$. Note that for any real numbers $p$ and $q$ we have
		\begin{equation}
			\binom{p}{q}=\frac{\Gamma(p+1)}{\Gamma(q+1)\,\Gamma(p-q+1)}\,.
		\end{equation}
		\item {\textbf{Polylogarithm function}}	\;\;\\\\
		The polylogarithm function is a special function of order $s$ and argument $z$ which is denoted as $\text{Li}_{s}(z)$. It is defined by a power series which is valid for arbitrary complex order $s$ and for all complex values of $z$ where $|z|<1$
		\begin{equation}
		\label{polylog}
		\text{Li}_{s}(z) = \sum_{k=1}^\infty \frac{z^k}{k^s}~.
		\end{equation}
		 By analytic continuation, the polylogarithm function can be extended to $\lvert z\rvert\ge 1$. For $s=1$ it is the usual known logarithm function 
		\begin{equation}
		\text{Li}_{1}(z) = -\text{ln} \left(1-z\right)~.
		\end{equation}
		 Also for $\mathfrak{Re}(s)>0$ and $\lvert z\rvert> 1$ its leading term is given by \cite{r11}
		\begin{equation}\label{eq18b}
			\text{Li}_s(z)\sim -\frac{\left[\ln\left(z\right)\right]^s}{\Gamma\left(s+1\right)}\,.
		\end{equation}

        \item {\textbf{Hypergeometric function expansions}}	\;\;\\\\
        The hypergeometric function accepts two different limits. One where $z \rightarrow 0$ which leads to a convergent series for the hypergeometric function and the other one is the limit $z \rightarrow 1$. For the second one the hypergeometric function is not necessarily convergent.    

Taylor series expansion for the Hypergeometric function as $z \rightarrow 0$ is:
\begin{align}\label{hypergeometric function}
_2F_1(a,b,c;z)&=\frac{\Gamma[c]}{\Gamma[b]}\sum_{n=0}^{\infty}\frac{\Gamma[n+b]}{\Gamma[n+c]}\frac{(a)_{n}}{n!}z^n~;\quad (a)_{n}=a(a+1)\dots (a+n-1)~,\nonumber \\
  _2F_1(a,b,c;z)&=1+\frac{ab}{c}z+\frac{a(a+1)b(b+1)}{c(c+1)}\frac{z^2}{2!}+\dots~.
\end{align}
For the second limit where $z \rightarrow 1$, if the parameters in the hypergeometric function satisfy Real$(c-a-b)>0$ then $_2F_1(a,b;c,Z)$ is absolutely convergence. So, we can it as linear combination of two other convergent hypergeometric functions with argument $1-z$
 \begin{align}
 \label{hypid}
    _2F_1(a,b;c,z)&=\frac{\Gamma[c]\Gamma[c-a-b]}{\Gamma[c-a]\Gamma[c-b]}~_2F_1(a,b;a+b+1-c,1-z)\nonumber\\
                   &+~\frac{\Gamma[c]\Gamma[a+b-c]}{\Gamma[a]\Gamma[b]}~(1-z)^{c-a-b}~_2F_1(c-a,c-b;1+c-a-b,1-z).
 \end{align}
	\end{enumerate}
	%

	\section{Numerical coefficients}\label{app2}
		Here is the list of all coefficients defined throughout the paper:
		\begin{equation}
		\label{ck}
		 c_k \equiv \frac{\sqrt{\pi}\Gamma\left(\frac{d(k+1)}{2(d-1)}\right)}{(1+kd)\Gamma\left(\frac{1+kd}{2(d-1)}\right)}~,
		 \end{equation}
		 %
and
\begin{equation}
\begin{aligned}
\label{as}
  &a_1 \equiv \frac{(2 c_0)^{d-1}}{2(2-d)},\\
  &a_2 \equiv \frac{1}{2d}(2 c_0)^{2(d-1)},\\
  &a_3 \equiv \frac{\left(1-d\right)}{d^2}(2c_0)^{3(d-1)},\\
  &a_4 \equiv \frac{\left(d-1\right)c_1}{4(2 c_0)^2},\\
  &a_5 \equiv \frac{\left(d-1\right)c_1(2 c_0)^{d-3}}{d},\\
  &a_6 \equiv \frac{(2c_0)^{2(d-1)}}{8d}-\frac{\left(4d^2-7d+3\right)c_1 (2{c_0})^{2(d-2)}}{d^2},\\
  &a_7 \equiv \frac{3 \left(d-1\right) c_2}{8\left(d+2\right)(2c_0)^{d+2}}-\frac{\left(d-1\right) {c_1}^2}{4(2 c_0)^{d+3}},\\
  &a_8 \equiv \frac{\left(2d^2-d+3\right) {c_1}^2}{2d(2c_0)^4}-\frac{3c_2}{4d(2c_0)^3},\\
  &a_9 \equiv \frac{(2 c_0)^{d-5}}{4 d^2}\bigg(4d\left(2d-1\right){c_0}^2 c_1 - 2\left(9d^3-36d^2+19d-12\right){c_1}^2-6\left(5d^2-2d+3\right)c_0 c_2 \bigg).
    \end{aligned}
    \end{equation}

		\vfill
	
\end{document}